\documentclass[useAMS,usenatbib]{mn2e}
\usepackage{graphicx}
\usepackage[T1]{fontenc}
\usepackage{aecompl}
\title[Intrinsic $\gamma$-ray luminosity, black hole mass, jet and accretion in Fermi blazars]{Intrinsic $\gamma$-ray luminosity, black hole mass, jet and accretion in Fermi blazars}
\author[D. R. Xiong, X. Zhang]{D. R. Xiong$^{1,3}$, X. Zhang$^{2}\thanks{E-mail: ynzx@yeah.net}$\\
$^{1}$National Astronomical Observatories/Yunnan Observatories, Chinese Academy of Sciences, Kunming 650011, China\\
$^{2}$Department of Physics, Yunnan Normal University, Kunming 650500, China\\
$^{3}$The Graduate School of Chinese Academy of Sciences, Beijing
100049, China}
\begin{document}

\pagerange{\pageref{firstpage}--\pageref{lastpage}} \pubyear{2002}

\maketitle

\label{firstpage}

\begin{abstract}
We have analyzed a large sample of clean blazars detected by Fermi
Large Area Telescope (LAT). Using literature and calculation, we
obtained intrinsic $\gamma$-ray luminosity excluding beaming effect,
black hole mass, broad-line luminosity (used as a proxy for disk
luminosity), jet kinetic power from ``cavity'' power and bulk
Lorentz factor for parsec-scale radio emission, and studied the
distributions of these parameters and relations between them. Our
main results are as follows. (i) After excluding beaming effect and
redshift effect, intrinsic $\gamma$-ray luminosity with broad line
luminosity, black hole mass and Eddington ratio have significant
correlations. Our results confirm the physical distinction between
BL Lacs and FSRQs. (ii) The correlation between broad line
luminosity and jet power is significant which supports that jet
power has a close link with accretion. Jet power depends on both the
Eddington ratio and black hole mass. We also obtain $LogL_{\rm
BLR}\sim(0.98\pm0.07)Log P_{\rm jet}$ for all blazars, which is
consistent with the theoretical predicted coefficient. These results
support that jets are powered by energy extraction from both
accretion and black hole spin (i.e., not by accretion only). (iii)
For almost all BL Lacs, $P_{\rm jet}>L_{\rm disk}$; for most of
FSRQs, $P_{\rm jet}<L_{\rm disk}$. The ``jet-dominance''
(parameterized as $\frac{P_{\rm jet}}{L_{\rm disk}}$) is mainly
controlled by the bolometric luminosity. Finally, the radiative
efficiency of $\gamma$-ray and properties of TeV blazars detected by
Fermi LAT were discussed.
\end{abstract}


\section{Introduction}

Blazars are the most extreme active galactic nuclei (AGN) pointing
their jets in the direction of the observer (Urry \& Padovani 1995),
and are the brightest and the most dominant population of AGN in the
$\gamma$-ray sky (Fichtel et al. 1994; Abdo et al. 2010a). Their
extremely observational properties are explained as being due to a
beaming effect. Due to a relativistic beaming effect, the emissions
from the jet are strongly boosted in the observer's frame (Urry \&
Padovani 1995). Blazars are often divided into two subclasses of BL
Lacertae objects (BL Lacs) and flat spectrum radio quasars (FSRQs).
FSRQs have strong emission lines, while BL Lac objects have only
very weak or non-existent emission lines. The classic division
between FSRQs and BL Lacs is mainly based on the equivalent width
(EW) of the emission lines. Objects with rest frame EW$>5$ {\AA} are
classified as FSRQs (e.g., Scarpa \& Falomo 1997; Urry \& Padovani
1995). Blandford \& Rees (1978) had originally suggested that the
absence of broad lines in BL Lacs was due to a very bright,
Doppler-boosted synchrotron continuum. On the other hand, EW greater
than 5 {\AA} may be the results of a particularly low state of the
beamed continuum in a source of intrinsically weak lines, and the
jet electromagnetic output is often dominated by the emission at
higher energies (Ghisellini et al. 2011; Sbarrato et al. 2012).
Therefore the EW alone is not a good indicator of the distinction
between the two classes of blazars. By studying the transition
between BL Lacs and FSRQs, Ghisellini et al. (2011) proposed a
physical distinction between the two classes of blazars, based on
the luminosity of the broad line region measured in Eddington units,
and the dividing line is of the order of
$L_{\rm{BLR}}/L_{\rm{Edd}}\sim 5\times10^{-4}$. The result also was
confirmed by Sbarrato et al. (2012).

Many models have been proposed to explain the origin of the blazar
$\gamma$-ray emission, including synchrotron self-Compton (SSC)
(e.g., Maraschi et al. 1992), inverse Compton (IC) scattering on
photons produced by the accretion disk (Dermer et al. 1992; Zhang \&
Cheng 1997), scattered by ambient material, or reprocessed by the
broad line clouds (Sikora et al. 1994; Xie et al. 1997), synchrotron
emission by ultrarelativistic electrons and positrons (e.g.,
Ghisellini et al. 1993; Cheng et al. 1993), and electromagnetic
cascade by collision of ultrarelativistic nucleons (e.g., Mannheim
\& Biermann 1992; Mannheim 1993; Cheng \& Ding 1994). The most
popular opinion is that $\gamma$-ray of powerful blazars is produced
within the BLR via EC (e.g., Sikora et al. 1994; Ghisellini \& Madau
1996) and $\gamma$-ray of low power blazars is SSC (Maraschi et al.
1992). Since the launch of the Fermi satellite, we have entered in a
new era of blazars research (Abdo et al. 2009, 2010a, 2010b). Up to
now, the Large Area Telescope (LAT) has detected hundreds of blazars
because it has about 20 folds better sensitivity than its
predecessor EGRET in the 0.1-100 GeV energy rang. The dramatically
improved $\gamma$-ray data from Fermi LAT has opened up the
possibility of testing results obtained from the EGRET era regarding
the origin of $\gamma$-rays. Sbarrato et al. (2012) found a good
correlation between the luminosity of the broad lines and the
$\gamma$-ray luminosity. But the $\gamma$-ray luminosity in their
sample did not consider the beaming effect; the sample studied in
Sbarrato et al. (2012) was limited; most of BL Lacs from Sbarrato et
al. (2012) sample have been selected to be at $z<0.4$.

The radiation observed from blazars is dominated by the emission
from relativistic jets which transport energy and momentum to large
scales (Blandford \& Rees 1978). However, jet formation remains one
of the unsolved fundamental problems in astrophysics (e.g., Meier et
al. 2001). Many models have been proposed to explain the origin of
jets. In current theoretical models of the formation of jet, power
is generated via accretion and extraction of rotational energy of
disc/black hole (Blandford \& Znajek 1977; Blandford \& Payne 1982),
and then converted into the kinetic power of the jet. In both
scenarios the magnetic field plays a major role in channeling power
from the BH or the disk into the jet; in both cases it should be
sustained by matter accreting onto BH, leading one to expect a
relation between accretion and jet power (Maraschi \& Tavecchio
2003). The jet-disk connection has been extensively explored by many
authors and in different ways (e.g., Rawlings \& Saunders 1991;
Falcke \& Biermann 1995; Serjeant et al. 1998; Cao \& Jiang 1999;
Wang et al. 2004; Liu et al. 2006; Xie et al. 2007; Gu et al. 2009;
Ghisellini et al. 2009a, 2009b, 2010, 2011; Sbarrato et al. 2012).
More and more evidences show that the jet power of extragalactic
radio loud sources is of the same order (or slightly larger than) of
the disk luminosity (e.g., Rawlings \& Saunders 1991; Ghisellini et
al. 2009a, 2009b). On the larger scale (radio-lobe size), one find
out the minimum jet power needed to sustain the radio-lobe emission
via considering minimum energy and the lifetime of radio lobes
(Rawlings \& Saunders 1991). At smaller, but still very large, jet
scales (on kpc to Mpc), one can use the recently discovered X-ray
emission from resolved knots, to model it and to infer the total
number of leptons needed to produce the observed radiation. Assuming
a proton per emitting lepton, Tavecchio et al. (2000), Celotti,
Ghisellini \& Chiaberge (2001), Tavecchio et al. (2004), Sambruna et
al. (2006) and Tavecchio et al. (2007) derived jet powers. At the
VLBI scale (pc or tens of pc), one takes advantage of the resolving
power of VLBI to measure the size of the synchrotron emitting region
(Celotti \& Fabian 1993). And this gives an estimate of the jet
power. A recent technique makes use of the cavities or bubbles in
the X-ray emitting intra-cluster medium of cluster of galaxies, and
measures the energy required to inflate such bubbles. Assuming that
this energy is furnished by the jet, one can calculate the
associated jet power (e.g., Allen et al. 2006; Balmaverde et al.
2008; Cavagnolo et al. 2010).

Blazars detected in the TeV or very high energy regime (VHE; E$>$100
GeV) are still a small group but their number is rapidly increasing
and are intensively studied (e.g., De Angelis et al. 2008), because
they are good laboratories to investigate particle acceleration and
cooling and to indirectly probe the extragalactic background light
(EBL, e.g., Stecker et al. 1992; Stanev \& Franceschini 1998; Mazin
\& Raue 2007; Tavecchio et al. 2010, 2011). From TeVCat
catalog\footnote{http://tevcat.uchicago.edu}, the majority of the
detected TeV blazars belong to high-frequency peaked BL Lacs (HBLs).
There are three FSRQs detected in the TeV band (TFSRQs; 4C +21.35,
PKS 1510-089, 3C 279). Most of them have recently been detected also
at MeV-GeV energies by Fermi Large Area Telescope (LAT) (Abdo et al.
2010a, 2012). The TeV observations have shown dramatic variability
in some TeV blazars, which suggests extremely small emitting volumes
and/or time compression by large relativistic Doppler factors
(Aharonian et al. 2007; Wagner et al. 2007). It is generally
accepted that SSC model can account for the observed SED of TeV BL
Lacs (TBL Lacs). However this may be an over-simplification, since,
besides the jet region containing the energetic electrons
responsible for the high energy emission, other sites, both in the
jet and externally to it (e.g., a molecular torus, a thin scattering
plasma surrounding the jet, or the walls of the jet itself) may be
important in producing the soft seed photons to be scattered at high
energies (e.g., Costamante \& Ghisellini 2002).

In this paper, through constructing a large sample of clean Fermi
blazars and removing beaming effect, we studied the correlations
between intrinsic $\gamma$-ray luminosity and black hole mass,
between intrinsic $\gamma$-ray luminosity and Eddington ratio, and
revisited the correlation between the luminosity of the broad lines
and the intrinsic $\gamma$-ray luminosity and physical distinction
between the two classes of blazars. we estimated the jet kinetic
power for the Fermi blazars from Nemmen et al. (2012) to study the
jet-disk connection and properties of Fermi blazars detected in TeV
band.

The paper is structured as follows: in Sect. 2, we present the
samples; the results are presented in Sect. 3 and discussions are in
Sect. 4; our conclusions are presented in Sect. 5. The cosmological
parameters $H_{\rm 0}=70~ {\rm km~s^{-1}~Mpc^{-1}}$, $\Omega_{\rm
m}=0.3$, and $\Omega_{\rm \Lambda}=0.7$ have been adopted in this
work.

\section{The samples}

We tried to select the largest group of clean blazars detected by
Fermi LAT with reliable broad line luminosity, $\gamma$-ray
luminosity, redshift, black hole mass and jet kinetic power. For the
aim, we collected many very large samples of blazars to get broad
line data and black hole mass and cross-correlated these sample with
clean blazars detected by Fermi LAT. Firstly, we considered the
following samples of blazars to get the broad line data: Cao \&
Jiang (1999), Wang et al. (2004), Liu et al. (2006), Sbarrato et al.
(2012), Chai et al. (2012), Shen et al. (2011), Shaw et al. (2012).
We cross-correlated these sample with clean blazars detected by
Fermi LAT in two years of scientific operation (Abdo et al. 2012,
2FGL; Ackermann et al. 2011a, 2LAC). Secondly, we considered the
following samples of blazars to get black hole: Woo \& Urry (2002),
Xie et al. (2004), Liu et al. (2006), Zhou \& Cao (2009), Zhang et
al. (2012), Sbarrato et al. (2012), Chai et al. (2012), Leon-Tavares
et al. (2011a), Shen et al. (2011), Shaw et al. (2012). At last, we
cross-correlated these Fermi blazars with sample of Nemmen et al.
(2012) to get jet kinetic power and beaming factor. In total, we
have a sample containing 248 clean Fermi blazars (191 FSRQs and 57
BL Lacs), including 20 TBL Lacs and 3 TFSRQs.

\subsection{Intrinsic $\gamma$-ray luminosity}

The Fermi satellite is detecting $\gamma$-ray emission from a large
number of blazars. The second catalog of active galactic nuclei
(2LAC) detected by the Fermi Large Area Telescope (LAT) in two years
of scientific operation includes 1017 $\gamma$-ray sources located
at high Galactic latitudes ($|b|>10^\circ$) (Ackermann et al.
2011a). These $\gamma$-ray sources are detected with a test
statistic (TS) greater than 25 and associated statistically with
AGNs. However, some of these sources are affected by analysis issues
and associated with multiple AGNs. Consequently, the clean sample
includes 886 AGNs, comprising 395 BL Lacs, 310 FSRQs, 157 candidate
blazars of unknown type, eight misaligned AGNs, four narrow-line
Seyfert 1 (NLS1s), 10 AGNs of other types and two starburst
galaxies. Source detection is based on the average flux over the
24-month period and flux measurements are included in 5 energy bands
(Abdo et al. 2012). We also note that 56\% of the BL Lacs have no
measured redshifts.

Nemmen et al. (2012) established a physical analogy between AGNs and
gamma-ray bursts (GRBs). A key point in their work was that
$\gamma$-ray luminosity of blazars and GRBs considered beaming
effect. They computed the intrinsic $\gamma$-ray luminosity $L$ of
blazars by correcting observation $\gamma$-ray luminosity $L^{obs}$
for the beaming factor $f_b$, such that $L=f_bL^{obs}$. For blazars,
$f_b$ was estimated as $1-cos(1/\Gamma$) where $\Gamma$ was the bulk
Lorentz factor of the flow, since jet opening angle $\theta_j$ of
AGNs obey $\theta_j<1/\Gamma$ (Jorstad et al. 2005; Pushkarev et al.
2009). Using VLBI and VLBA, Hovatta et al. (2009) and Pushkarev et
al. (2009) calculated the variability Lorentz factors $\Gamma_{\rm
var}$. The bulk Lorentz factors of Nemmen et al. (2012) were from
the results of Hovatta et al. (2009) and Pushkarev et al. (2009).
Because $\theta$ was not available for the whole blazar sample, they
used the power-law fit of
$f_b\approx5\times10^{-4}(L^{obs}_{49})^{-0.39\pm0.15}$ as an
estimator for $f_b$.

Nemmen et al. (2012) computed the K-corrected $\gamma$-ray
luminosity at 0.1-100 GeV, uncertainty and the intrinsic
$\gamma$-ray luminosity excluding beaming effect. The intrinsic
$\gamma$-ray luminosity of some blazars from our sample are obtained
from Nemmen et al. (2012). If intrinsic $\gamma$-ray luminosity in
our sample are not obtained from Nemmen et al. (2012), we follow a
procedure similar of Nemmen et al. (2012) to calculate intrinsic
$\gamma$-ray luminosity and use average uncertainty with 0.6 dex for
them (we also use the power-law fit of
$f_b\approx5\times10^{-4}(L^{obs}_{49})^{-0.39\pm0.15}$ as an
estimator for $f_b$).

\subsection{Broad line luminosity}

The BLR luminosity given in Celotti, Padovani \& Ghisellini (1997)
were derived by scaling several strong emission lines to the quasar
template spectrum of Francis et al. (1991), and used $\rm{Ly}\alpha$
as a reference. Sbarrato et al. (2012) had taken the luminosity of
emission lines of the blazars in SDSS DR7 Quasar sample. For
calculating the total luminosity of the broad lines, they had
followed Celotti, Padovani \& Ghisellini (1997). Specifically they
set $\rm{Ly}\alpha$ flux contribution to 100, the relative weight of
$\rm{H}\alpha$, $\rm{H}\beta$, MgII and CIV lines respectively to
77, 22, 34 and 63. The total broad line flux was fixed at 555.76.
Their broad-line luminosity had been derived using these
proportions. When more than one line was present, they calculated
the simple average of broad-line luminosity estimated from each
line. The rest of authors in our sample also adopted the method
proposed by Celotti, Padovani \& Ghisellini (1997) and similar
processes to gain broad-line luminosity.

Shaw et al. (2012) reported on optical spectroscopy of 229 blazars
in the Fermi 1LAC sample and Shen et al. (2011) had spectrally
analyzed SDSS DR7 Quasar sample. The luminosity of emission lines of
our some blazars are from Shaw et al. (2012) and Shen et al. (2011),
and we use similar method of Sbarrato et al. (2012) to calculate
broad-line luminosity. However, we find that some objects of
broad-line luminosity are distinct from different samples with our
results. The possible reasons are that using lines to calculate
broad-line luminosity is different and variability also can cause
the difference of them. In these sources, we use average broad-line
luminosity instead.

\subsection{Black hole mass}

The traditional virial black hole masse is estimated by using an
empirical relation between BLR size and ionizing luminosity together
with measured broad-line widths assuming the BLR clouds being
gravitationally bound by the central black hole. For most of FSRQs
in our sample, the black hole mass is estimated by traditional
virial method (Woo \& Urry 2002; Wang et al. 2004; Liu et al. 2006;
Sbarrato et al. 2012; Chai et al. 2012; Shaw et al. 2012; Shen et
al. 2011). When virial black hole masses are attained from different
lines, we simply average black hole masses from different lines. In
Shen et al. (2011), we get black hole masses from Vestergaard \&
Peterson (2006) for $\rm{H}\beta$ and CIV, and Vestergaard \& Osmer
(2009) for MgII. For some sources, especially BL Lac objects, the
black hole masses can be estimated from the properties of their host
galaxies with either $M_{\rm BH}-\sigma$ or $M_{\rm BH}-L$
relations, where $\sigma$ and $L$ are the stellar velocity
dispersion and the bulge luminosity of the host galaxies (Woo \&
Urry 2002; Zhou \& Cao 2009; Zhang et al. 2012; Sbarrato et al.
2012; Chai et al. 2012; Leon-Tavares et al. 2011a). For a few
sources, the black hole masses are estimated from variation
timescale (Xie et al. 1991, 2004). For same blazars, when more than
one black hole masses are got, we use average black hole mass
instead.

\subsection{jet kinetic power}

Cavagnolo et al. (2010) searched for X-ray cavities in different
systems including giant elliptical galaxies and cD galaxies and
estimated the jet power required to inflate these cavities or
bubbles, obtaining a tight correlation between the ``cavity'' power
and the radio luminosity
\begin{equation}
P_{\rm cav}\approx5.8\times10^{43}(\frac{P_{\rm radio}}{10^{40}{\rm
erg~s^{-1}}})^{0.7} {\rm erg~s^{-1}},
\end{equation}
which is continuous over $\sim6-8$ decades in $P_{\rm jet}$ and
$P_{\rm radio}$ with a scatter of $\approx0.7$ dex and $P_{\rm
jet}=P_{\rm cav}$. While this method is limited to a small number of
sources at present, the $P_{\rm jet}$ and $P_{\rm radio}$ relation
covers over $\sim6-8$ orders of magnitude in jet power, including
both FR I and FR II sources. Making use of the correlation between
$P_{\rm jet}$ and $P_{\rm radio}$ from Cavagnolo et al. (2010),
Meyer et al. (2011) chose the low-frequency extended luminosity at
300 MHz as an estimator of the jet power for blazars. Following
Meyer et al. (2011), Nemmen et al. (2012) estimated the jet kinetic
power for a large sample of Fermi blazars and obtained the relation
between intrinsic $\gamma$-ray luminosity and the kinetic power. The
best-fit parameters obtained from Nemmen et al. (2012) were $\alpha
= 0.98 \pm 0.02$ and $\beta = 1.6 \pm 0.9$ where $Log P_{jet} =
\alpha Log L_\gamma^{int} + \beta$. The scatter about the best-fit
is 0.64 dex. Our sample's jet powers are got from Nemmen et al.
(2012). When jet powers of blazars from our sample are not directly
got from Nemmen et al. (2012), we use the relation between intrinsic
$\gamma$-ray luminosity and the kinetic power from Nemmen et al.
(2012) to estimate the jet kinetic power. The uncertainty in $P_{\rm
jet}$ is dominated by the scatter in the correlation of Cavagnolo et
al. (2010) and corresponds to 0.7 dex.

The relevant data are listed in Table 1 with the following headings:
column (1), name of the Fermi LAT catalog; column (2), other name;
column (3) is right ascension (the first entry) and declination (the
second entry); column (4), classification of source$-$the first
entry: bzb=BL Lacs, bzq=FSRQs, tbzb=BL Lacs detected in the TeV or
very high energy regime and tbzq=FSRQs detected in the TeV or very
high energy regime; the second entry: classification of SED proposed
by Abdo et al. (2010c) and Ackermann et al. (2011b) (the synchrotron
peak frequency $\nu^s_{\rm{peak}}<10^{14}Hz$ for
low-synchrotron-peaked blazar LSP,
$10^{14}Hz<\nu^s_{\rm{peak}}<10^{15}Hz$ for
intermediate-synchrotron-peaked blazar ISP,
$10^{15}Hz<\nu^s_{\rm{peak}}$ for high-synchrotron-peaked blazar
HSP); column (5), redshift; column (6), logarithm of intrinsic
$\gamma$-ray luminosity excluding beaming effect in units of
${\rm{erg~s^{-1}}}$ and uncertainty; column (7), logarithm of black
hole mass in units of $M_{\rm \odot}$ and references; column (8),
logarithm of jet kinetic power with 0.7 dex uncertainty in units of
${\rm{erg~s^{-1}}}$ and logarithm of beaming factor; column (9),
logarithm of broad-line luminosity in units of ${\rm{erg~s^{-1}}}$
and references.

\section{The results}
\subsection{The distributions}

The redshift distributions of the various classes are shown in Fig.
1. They are very similar with complete 2LAC sample. The redshift
distributions for all blazars are $0<z<3.1$ and mean value is
$1.006\pm0.04$. Mean values for FSRQs and BL Lacs are $1.17\pm0.05$
and $0.45\pm0.05$ respectively. Compared with normal GeV BL Lacs
(NBL Lacs, $z=0.62\pm0.07$), blazars detected in the TeV or very
high energy regime (TBL Lacs) have much smaller mean redshift
($0.13\pm0.02$). The mean redshift of TeV FSRQs (TFSRQs) is
$0.44\pm0.09$. Through nonparametric Kolmogorov-Smirnov (KS) test,
we get that the redshift distributions between TBL Lacs and NBL Lacs
are significant difference (chance probability $P<0.0001$,
significant difference with $P<0.05$ confidence level); the redshift
distributions among HBLs (HSP BL Lacs), IBLs (ISP BL Lacs) and LBLs
(LSP BL Lacs) are significant difference ($P=0.003, P<0.0001,
P=0.006$).

The black hole mass distributions of the various classes are shown
in Fig. 2. The black hole mass distributions for all blazars mainly
are $10^{7.5}-10^{10}$ $M_{\rm \odot}$ and mean value is
$10^{8.5\pm0.03}$ $M_{\rm \odot}$. Mean values for FSRQs and BL Lacs
are $10^{8.55\pm0.04}$ $M_{\rm \odot}$ and $10^{8.34\pm0.06}$
$M_{\rm \odot}$ respectively. There are similar mean black hole
masses for NBL Lacs and TBL Lacs ($10^{8.36\pm0.08}$ $M_{\rm
\odot}$, $10^{8.31\pm0.09}$ $M_{\rm \odot}$). The mean black hole
mass of TFSRQs is $10^{8.53\pm0.11}$ $M_{\rm \odot}$. The black hole
mass distributions between TBL Lacs and NBL Lacs are not significant
difference ($P=0.345$). The black hole mass distributions among
HBLs, IBLs and LBLs are not significant difference ($P=0.4, 0.77,
0.1$). We note three blazars with a very low mass of the central
black hole (J0217.5-0813: $10^{6.53\pm0.61} M_{\rm \odot}$;
J0430.4-2507: $10^{6.51\pm0.77} M_{\rm \odot}$; J1954.6-1122:
$10^{6.73\pm0.39} M_{\rm \odot}$). The black hole mass of the three
blazars are directly from Shaw et al. (2012) in which the black hole
masses were estimated by traditional virial method. Shaw et al.
(2012) have urged caution in black hole mass of their blazars sample
because of non-thermal dominance. We also note that the FWHM of MgII
for the three blazars are small ($1200\pm400, 1200\pm200,
1500\pm600~{\rm km~s^{-1}}$). So the black hole masses of the three
blazars require further study. If the black hole masses of the three
blazars are indeed small, then it is very important for studying jet
of AGN, since the only known jetted AGN with low masses are
narrow-line Seyfert 1 galaxies.

The jet kinetic power distributions of the various classes are shown
in Fig. 3. The jet kinetic power distributions for all blazars are
$10^{42}-10^{47}\rm{erg~s^{-1}}$ and mean value is
$10^{45.08\pm0.04} \rm{erg~s^{-1}}$. Mean values for FSRQs and BL
Lacs are $10^{45.28\pm0.04} \rm{erg~s^{-1}}$ and $10^{44.4\pm0.11}
\rm{erg~s^{-1}}$ respectively. Compared with NBL Lacs
($10^{44.72\pm0.11} \rm{erg~s^{-1}}$), TBL Lacs have much smaller
mean jet kinetic power ($10^{43.80\pm0.15} \rm{erg~s^{-1}}$). The
jet kinetic power distributions between TBL Lacs and NBL Lacs are
significant difference ($P<0.0001$). The mean jet kinetic power of
TFSRQs is $10^{45.35\pm0.23} \rm{erg~s^{-1}}$. The jet kinetic
powers among HBLs, IBLs and LBLs are significant difference
($P=0.026, P<0.0001, P=0.023$).

The intrinsic $\gamma$-ray luminosity distributions of the various
classes are shown in Fig. 4. The intrinsic $\gamma$-ray luminosity
distributions for all blazars are $10^{42}-10^{46.5}\rm{erg~s^{-1}}$
and mean value is $10^{44.37\pm0.05} \rm{erg~s^{-1}}$ (as a
comparison, the observational $\gamma$-ray luminosity distributions
for all blazars is $10^{43.11}-10^{49.12}\rm{erg~s^{-1}}$ and mean
value are $10^{46.85\pm0.07} \rm{erg~s^{-1}}$). Mean values for
FSRQs and BL Lacs are $10^{44.54\pm0.04} \rm{erg~s^{-1}}$ and
$10^{43.78\pm0.11} \rm{erg~s^{-1}}$ respectively. Compared with NBL
Lacs ($10^{44.05\pm0.13} \rm{erg~s^{-1}}$), TBL Lacs have smaller
mean intrinsic $\gamma$-ray luminosity ($10^{43.28\pm0.13}
\rm{erg~s^{-1}}$). The mean intrinsic $\gamma$-ray luminosity of
TFSRQs is $10^{44.37\pm0.24} \rm{erg~s^{-1}}$. The intrinsic
$\gamma$-ray luminosity distributions between TBL Lacs and NBL Lacs
are significant difference ($P=0.001$). The intrinsic $\gamma$-ray
luminosity distributions among HBLs, IBLs and LBLs are significant
difference ($P=0.033, P<0.0001, P=0.007$).

The $\gamma$-ray photon index distributions of the various classes
are shown in Fig. 5. The $\gamma$-ray photon index distributions for
all blazars are $1.3-3$ and mean value is $2.29\pm0.02$ (for FSRQs
the $\gamma$-ray photon index distribution is $1.9-3$ and $1.3-2.5$
for BL Lacs). Mean values for FSRQs and BL Lacs are $2.37\pm0.02$
and $2.02\pm0.03$ respectively. From complete 2LAC clean sample with
$F[E>100{\rm MeV}]>1.5\times10^8~{\rm Ph~cm^{-2}s^{-1}}$ (see their
Fig. 18), the $\gamma$-ray photon index distributions for FSRQs and
BL Lacs are $1.9-3$ and $1.6-2.5$ respectively. The average photon
spectral indexes for FSRQs and BL Lacs are 2.42 and 2.06
respectively. In our sample, there are four BL Lacs in the $1.3-1.6$
interval because the four BL Lacs have $F[E>100{\rm
MeV}]<1.5\times10^8~{\rm Ph~cm^{-2}s^{-1}}$. To sum up, the
$\gamma$-ray photon index distributions of our sample are very
similar to complete 2LAC sample. Compared with NBL Lacs
($2.15\pm0.03$), TBL Lacs have much smaller mean $\gamma$-ray photon
index ($1.79\pm0.05$). The mean $\gamma$-ray photon index of TFSRQs
is $2.21\pm0.05$. The $\gamma$-ray photon index distributions
between TBL Lacs and NBL Lacs are significant difference
($P<0.0001$). The $\gamma$-ray photon index distributions between
HBLs and IBLs, between HBLs and IBLs are significant difference
(both $P<0.0001$) but not significant difference between IBLs and
LBLs ($P=0.64$).

The bulk Lorentz factor $\Gamma$ distributions of the various
classes are shown in Fig. 6 (for the blazars without direct
estimates of $\Gamma$, we estimate bulk Lorentz factor by the
relation $f_b=1-cos\frac{1}{\Gamma}$). The bulk Lorentz factor
distributions for almost all blazars are 0-30 and mean value is
$13.76\pm0.44$. Mean values for FSRQs and BL Lacs are $15.18\pm0.5$
and $9.03\pm0.65$ respectively. Compared with NBL Lacs
($10.52\pm0.86$), TBL Lacs have smaller mean bulk Lorentz factor
($6.27\pm0.58$). The mean bulk Lorentz factor of TFSRQs is
$28.87\pm8.13$ (their bulk Lorentz factors are directly from the
radio measurements). The bulk Lorentz factor distributions between
TBL Lacs and NBL Lacs are significant difference ($P=0.001$). The
bulk Lorentz factor distributions between HBLs and IBLs, between
HBLs and LBLs are significant difference ($P=0.004$, $P<0.0001$) but
not significant difference between IBLs and LBLs ($P=0.18$). We also
compare our bulk Lorentz factor from the radio measurements
($\Gamma_{\rm R}$) with bulk Lorentz factor calculated by modeling
the SED from Ghisellini et al. (2010) ($\Gamma_{\rm G}$). Apart from
our sample, we also include 10 blazars from Nemmen et al. (2012)
because they are included in both Ghisellini et al. (2010) and
Nemmen et al. (2012) (J0120.4-2700, J0449.4-4350, J1719.3+1744,
J0205.3-1657, J0217.9+0143, J0221.0+3555, J1332.0-0508,
J2056.2-4715, J2147.3+0930, J2203.4+1726). But the blazars are not
included in our sample because their black hole mass and BLR data
can not be obtained. The result is shown in Fig. 7. From Fig. 7, we
can see that for all HBLs, $\Gamma_{\rm R}<\Gamma_{\rm G}$; for most
of IBLs and LBLs, $\Gamma_{\rm R}<\Gamma_{\rm G}$; for most of
NFSRQs and TFSRQs, $\Gamma_{\rm R}>\Gamma_{\rm G}$. At the end of
this section, we cross check the sample of blazars in Meyer et al.
(2011) and Nemmen et al. (2012) with that of Ghisellini et al.
(2010), and compare the kinetic jet power as measured with the two
methods. The result is shown in Fig. 8. From Fig. 8, we can see that
on average, the jet power from Ghisellini et al. (2010) is slightly
larger than ``cavity'' power from Meyer et al. (2011) and Nemmen et
al. (2012). The difference can be due to the strong difference of
time scales.

\begin{figure}
\includegraphics[width=95mm, height=95mm]{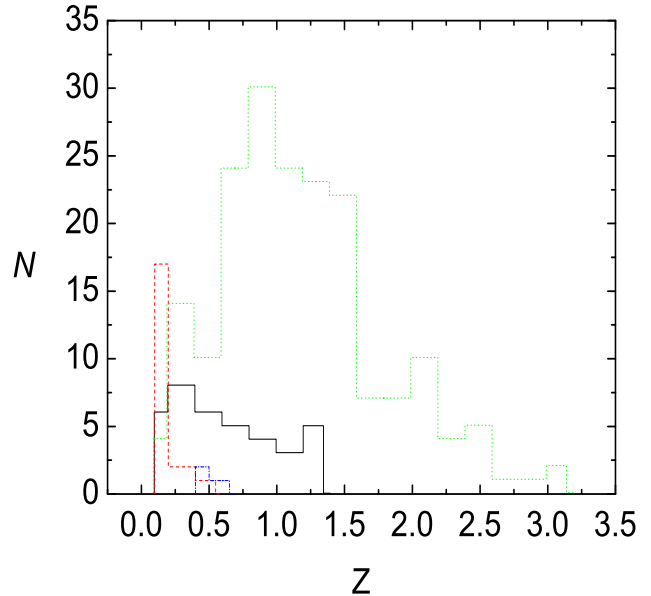}
\caption{Redshift distributions for normal GeV BL Lacs (NBL Lacs,
black continuous line), BL Lacs detected in the TeV or very high
energy regime (TBL Lacs, red dashed line), normal GeV FSRQs (NFSRQs,
green dotted line), FSRQs detected in the TeV or very high energy
regime (TFSRQs, blue dot--dashed line).} \label{figure 1}
\end{figure}

\begin{figure}
\includegraphics[width=95mm, height=87mm]{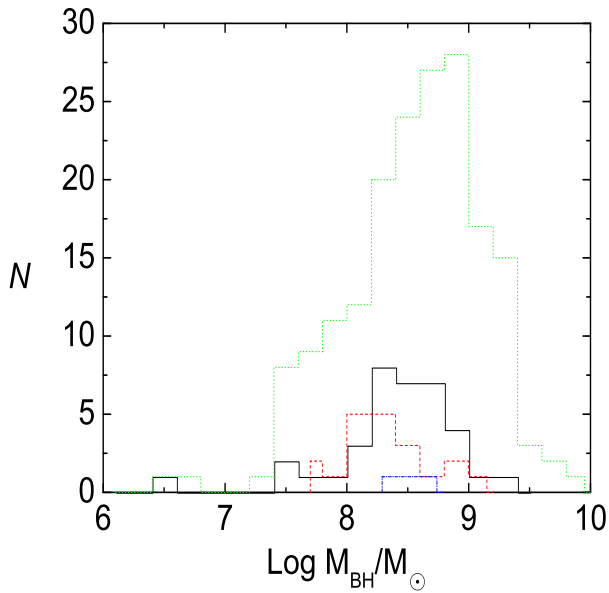}
\caption{Black hole mass distributions for NBL Lacs, TBL Lacs,
NFSRQs and TFSRQs. The meanings of different lines are as same as
Fig. 1.} \label{figure 2}
\end{figure}

\begin{figure}
\includegraphics[width=95mm, height=93mm]{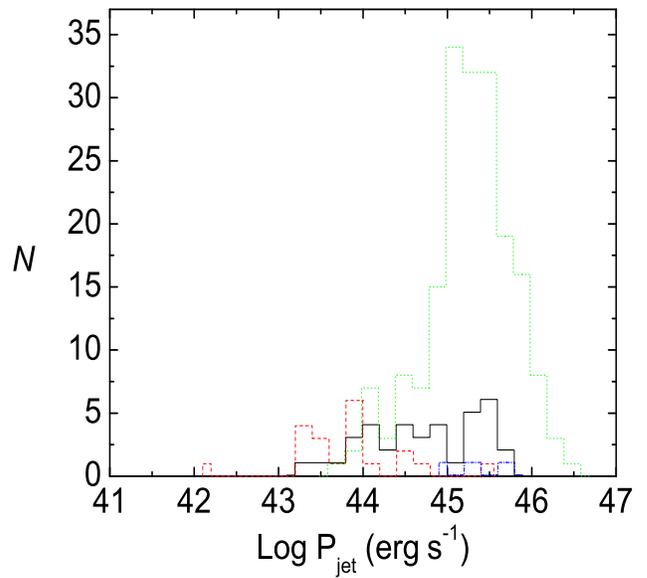}
\caption{Jet kinetic power distributions for NBL Lacs, TBL Lacs,
NFSRQs and TFSRQs. The meanings of different lines are as same as
Fig. 1.} \label{figure 1}
\end{figure}

\begin{figure}
\includegraphics[width=95mm, height=93mm]{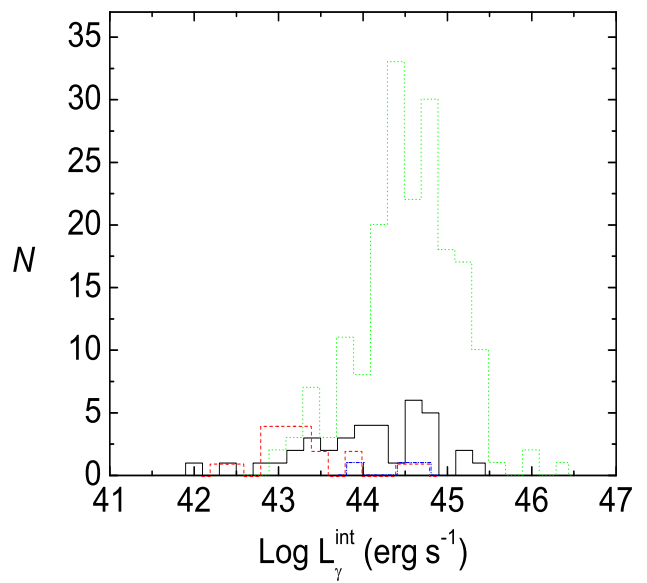}
\caption{Intrinsic $\gamma$-ray luminosity distributions for NBL
Lacs, TBL Lacs, NFSRQs and TFSRQs. The meanings of different lines
are as same as Fig. 1.} \label{figure 1}
\end{figure}

\begin{figure}
\includegraphics[width=95mm, height=87mm]{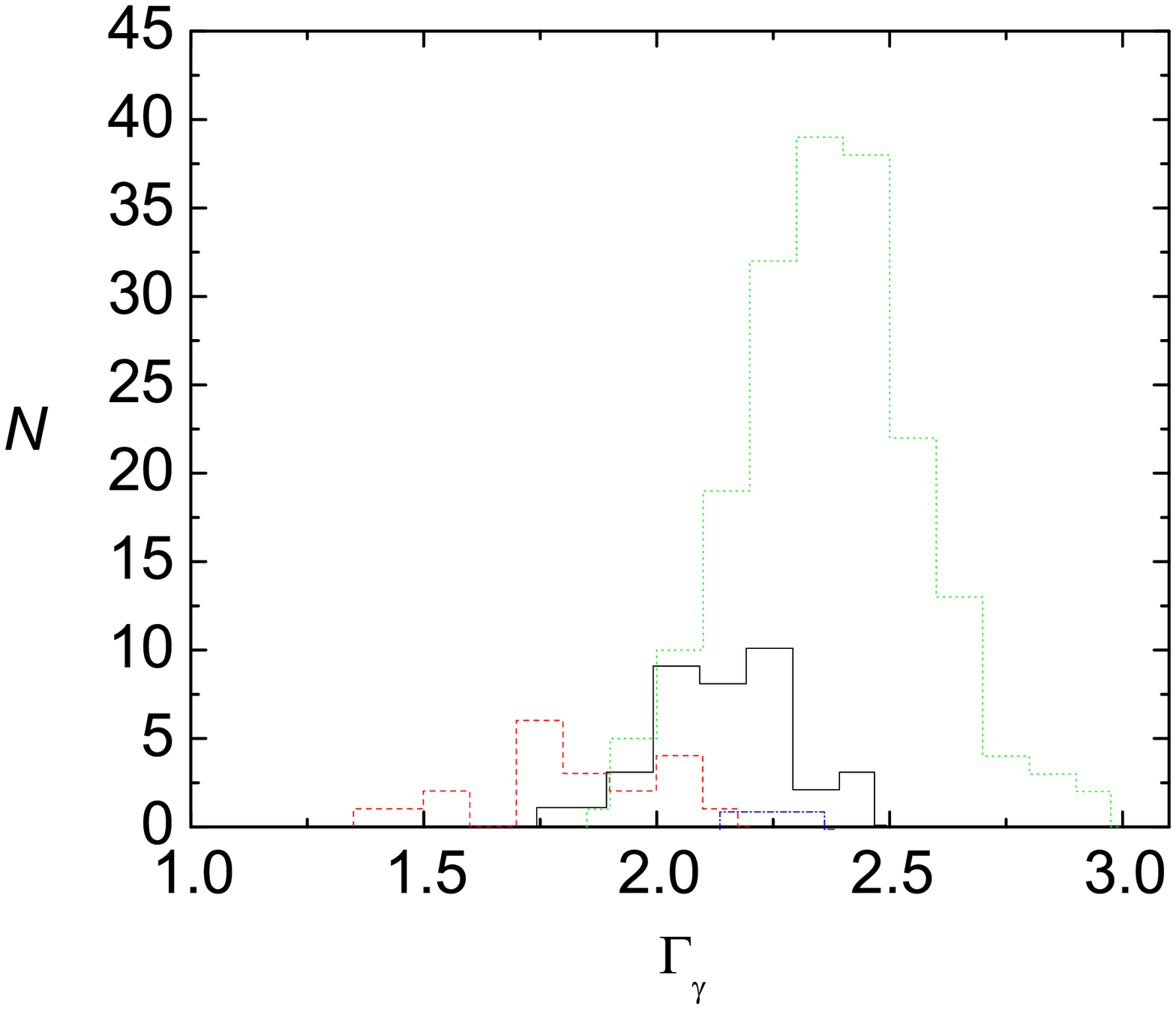}
\caption{$\gamma$-ray photon index distributions for NBL Lacs, TBL
Lacs, NFSRQs and TFSRQs. The meanings of different lines are as same
as Fig. 1.} \label{figure 1}
\end{figure}

\subsection{Intrinsic $\gamma$-ray luminosity vs black hole mass}

Figure 9 shows black hole mass as a function of intrinsic
$\gamma$-ray luminosity. Different symbols correspond to blazars
belonging to different classes. Pearson analysis is applied to
analyze the correlations between black hole mass and intrinsic
$\gamma$-ray luminosity for all blazars (Ackermann et al. 2011b;
Padovani 1992; Machalski \& Jamrozy 2006). The result shows that the
correlation between black hole mass and intrinsic $\gamma$-ray
luminosity is significant (number of points $N=239$, significance
level $P<0.0001$, coefficient of correlation $r=0.369$, significant
correlation $P<0.05$ confidence level). In view of that there are
correlations between black hole mass and redshift, between intrinsic
$\gamma$-ray luminosity and redshift, Pearson partial correlation
analysis excluding the dependence on the redshift is applied to
analyze the correlations between black hole mass and gamma-ray
luminosity. The result shows that the correlation between black hole
mass and intrinsic $\gamma$-ray luminosity is significant at 0.05
level when excluding redshift effect ($N=239$, $P=0.035$, $r=0.136$,
significant correlation $P<0.05$ confidence level). In Fig. 9, we
note that there are some objects out of main zone, which have black
hole mass above $10^{9.5} M_\odot$ and below $10^{7} M_\odot$. After
excluding these objects, we find a much better significance between
black hole mass and intrinsic $\gamma$-ray luminosity ($P=0.017$).

\subsection{intrinsic $\gamma$-ray luminosity vs broad line luminosity and Eddington ratio}

Figure 10 shows broad line luminosity as a function of intrinsic
$\gamma$-ray luminosity. Because there is also correlation between
broad line luminosity and redshift, Pearson partial correlation
analysis is applied to analyze the correlation. The result of
Pearson partial analysis shows that when excluding the dependence on
the redshift, there is still significant correlation between broad
line luminosity and intrinsic $\gamma$-ray luminosity ($N=217$,
$P<0.0001$, $r=0.321$). Figure 11 is Eddington ratio as a function
of $\gamma$-ray luminosity ($L_{\rm bol}/L_{\rm Edd}, L_{\rm
Edd}=1.3\times10^{38}(\frac{M}{M_\odot}){\rm erg~s^{-1}}, L_{\rm
bol}\approx10L_{\rm BLR}$ from Netzer (1990)). The result of Pearson
partial analysis also shows that after excluding the dependence on
the redshift, there is still significant correlation between
Eddington ratio and $\gamma$-ray luminosity ($N=208$, $P=0.001$,
$r=0.23$).

\begin{figure}
\includegraphics[width=95mm, height=87mm]{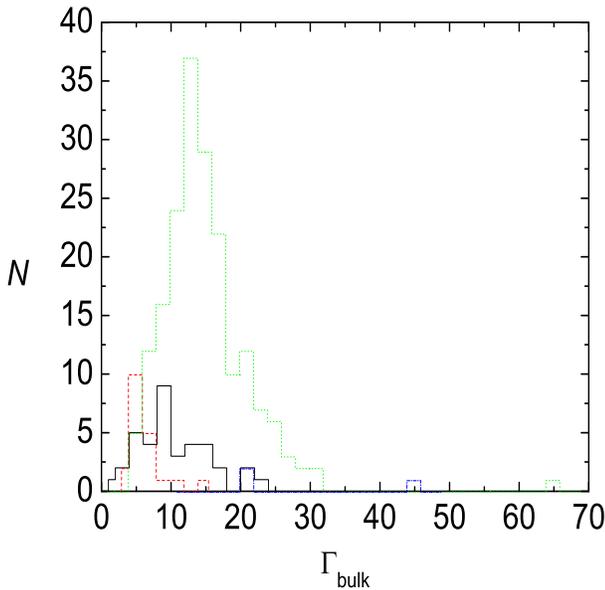}
\caption{Bulk Lorentz factor distributions for NBL Lacs, TBL Lacs,
NFSRQs and TFSRQs. The meanings of different lines are as same as
Fig. 1.} \label{figure 6}
\end{figure}

\begin{figure}
\includegraphics[width=95mm, height=87mm]{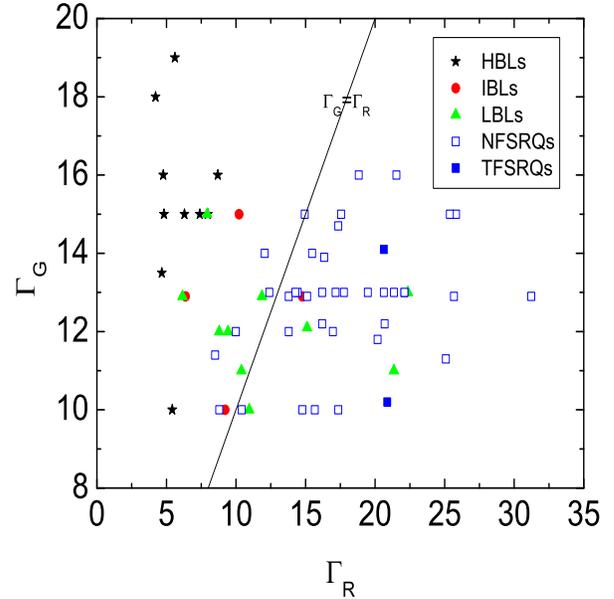}
\caption{Bulk Lorentz factor from radio measurements and bulk
Lorentz factor calculated by modeling the SED from Ghisellini et al.
(2010). The black line stands for that bulk Lorentz factors
calculated by the two methods are same. TFSRQs: blue filled squares;
NFSRQs: blue empty squares; LBLs: green filled triangles; IBLs: red
filled circles; HBLs: black stars.} \label{figure 6}
\end{figure}

\begin{figure}
\includegraphics[width=95mm, height=87mm]{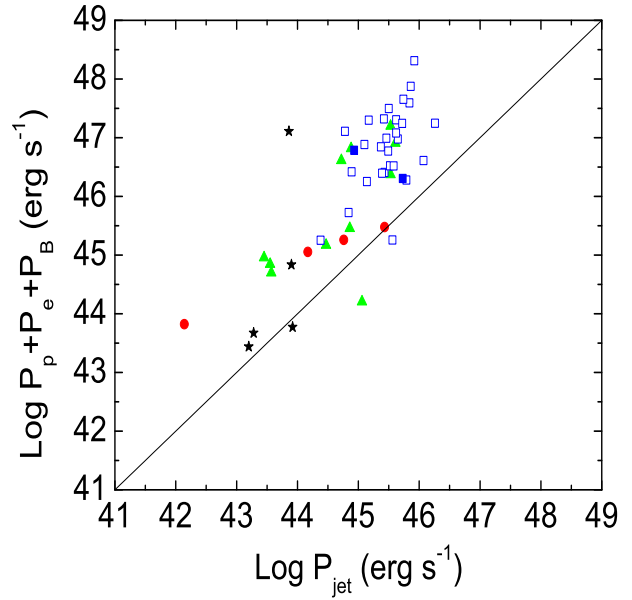}
\caption{Jet kinetic power based on X-ray cavities and jet power
calculated by modeling the SED from Ghisellini et al. (2010) (the
jet power $P_{\rm jet}=P_{\rm p}+P_{\rm e}+P_{\rm B}$, cold protons
$P_{\rm p}$, emitting electrons $P_{\rm e}$, Poynting flux $P_{\rm
B}$). The meanings of different symbols are as same as Fig. 7.}
\label{figure 6}
\end{figure}

\begin{figure}
\includegraphics[width=95mm, height=95mm]{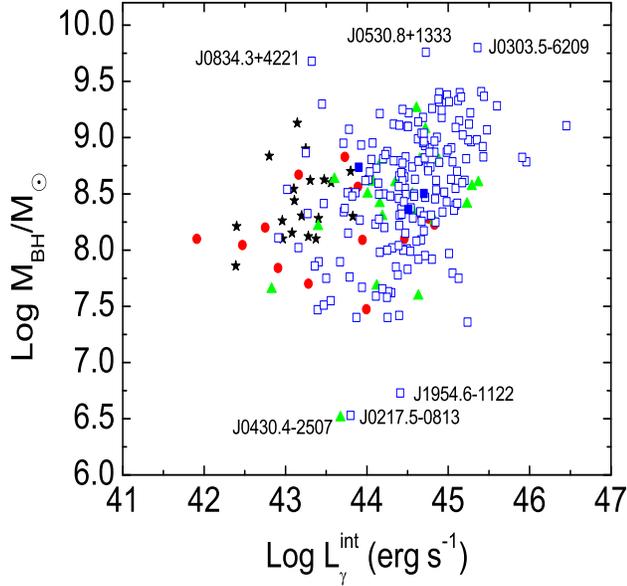}
\caption{Black hole mass as a function of intrinsic $\gamma$-ray
luminosity of various classes. The average uncertainty in intrinsic
$\gamma$-ray luminosity with and without direct estimates of
$\Gamma$ are 0.3 dex and 0.6 dex respectively. The meanings of
different symbols are as same as Fig. 7.} \label{figure 6}
\end{figure}

\begin{figure}
\includegraphics[width=95mm, height=95mm]{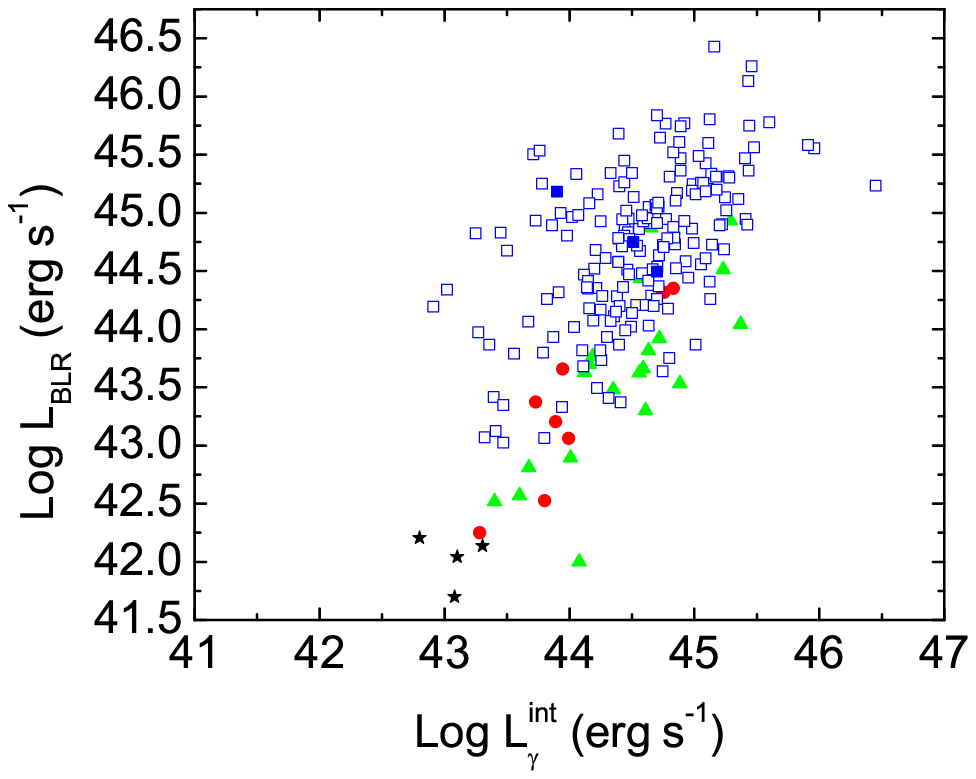}
\caption{Broad line luminosity as a function of intrinsic
$\gamma$-ray luminosity. The average uncertainty in intrinsic
$\gamma$-ray luminosity with and without direct estimates of
$\Gamma$ are 0.3 dex and 0.6 dex respectively. The meanings of
different symbols are as same as Fig. 7.} \label{figure 7}
\end{figure}

\begin{figure}
\includegraphics[width=95mm, height=95mm]{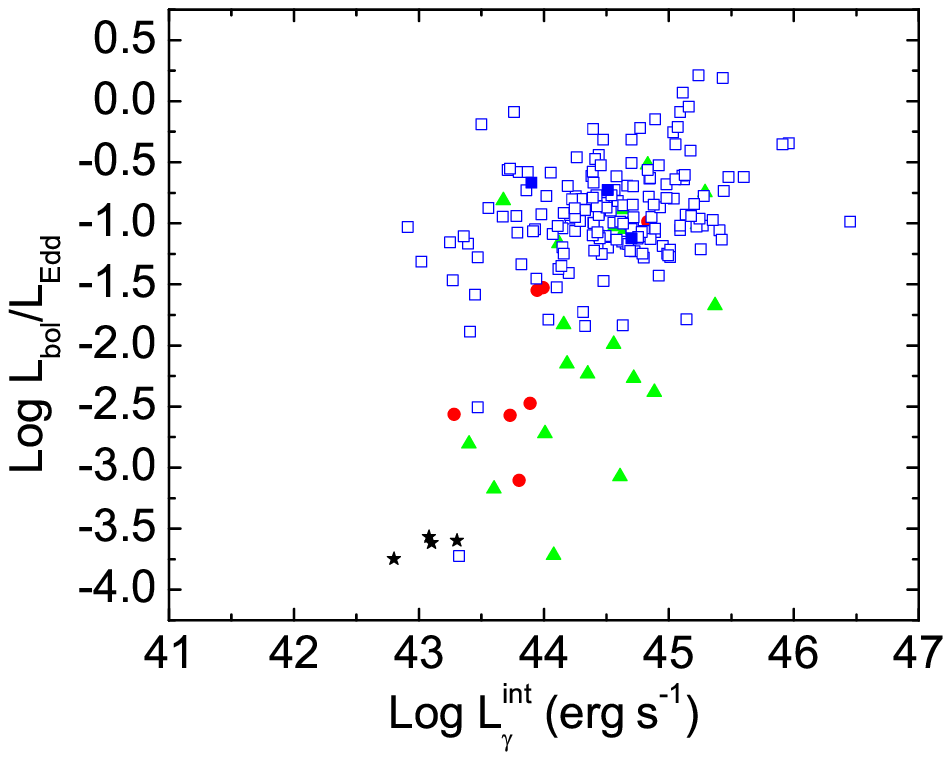}
\caption{Eddington ratio as a function of intrinsic $\gamma$-ray
luminosity. The average uncertainty in intrinsic $\gamma$-ray
luminosity with and without direct estimates of $\Gamma$ are 0.3 dex
and 0.6 dex respectively. The meanings of different symbols are as
same as Fig. 7.} \label{figure 8}
\end{figure}

\begin{figure}
\includegraphics[width=95mm, height=150mm]{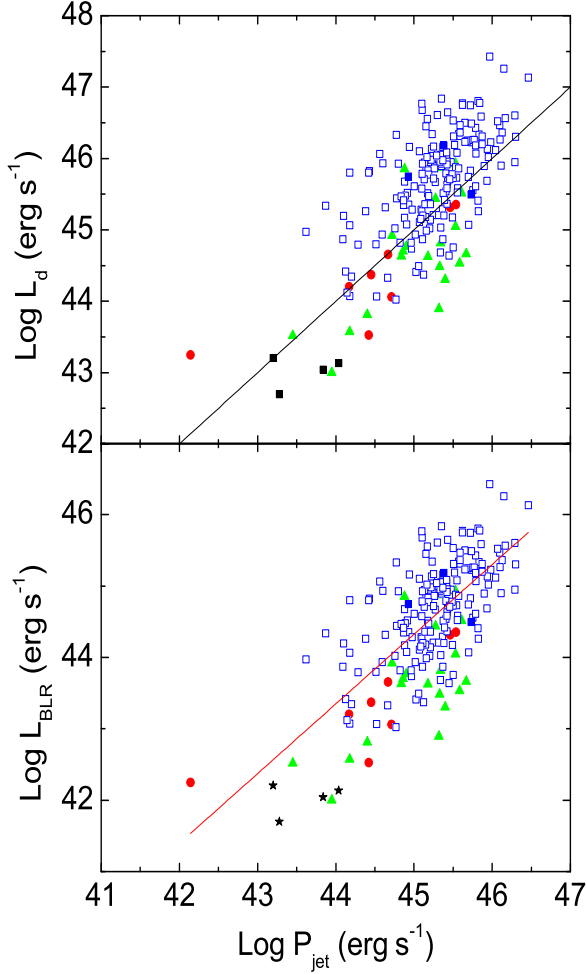}
\caption{Broad line luminosity as a function of jet power (bottom
panel) and disk luminosity as a function of jet power (top panel).
The red line is result of linear regression and black lines is
$L_{\rm d}=P_{\rm jet}$. The uncertainty of jet kinetic power is 0.7
dex. The meanings of different symbols are as same as Fig. 7.}
\label{figure 9}
\end{figure}

\begin{figure}
\includegraphics[width=95mm, height=95mm]{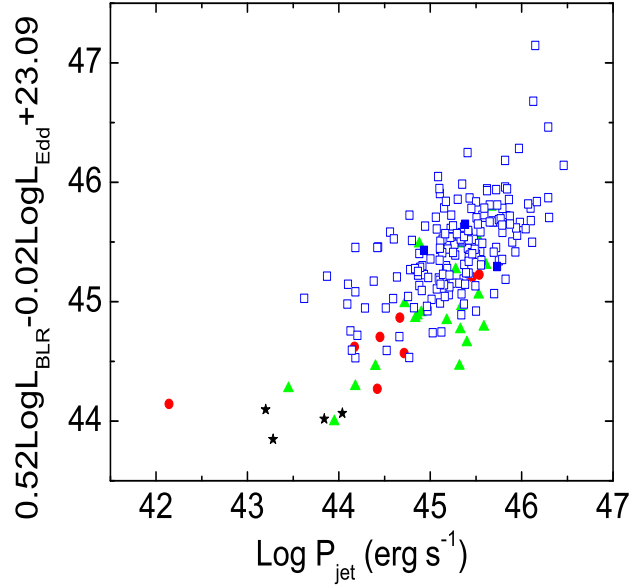}
\caption{Both broad line luminosity and Eddington luminosity as a
function of jet power. The uncertainty of jet kinetic power is 0.7
dex. The meanings of different symbols are as same as Fig. 7.}
\label{figure 10}
\end{figure}

\begin{figure}
\includegraphics[width=95mm, height=95mm]{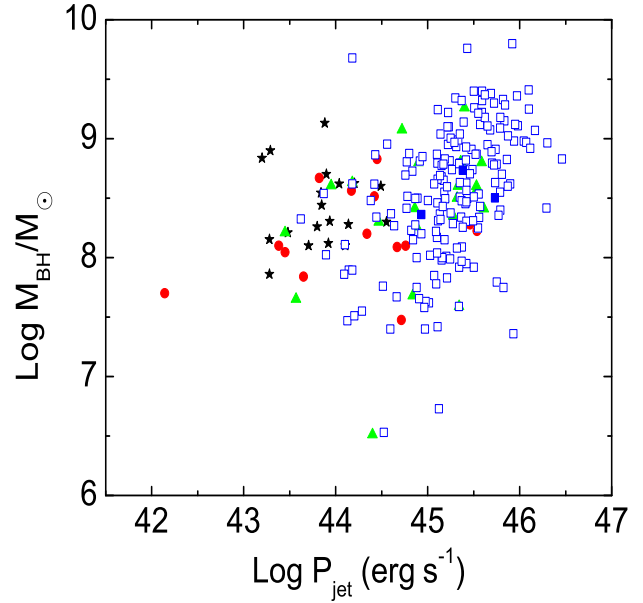}
\caption{The black hole mass as a function of jet power. The
uncertainty of jet kinetic power is 0.7 dex. The meanings of
different symbols are as same as Fig. 7.} \label{figure 10}
\end{figure}

\begin{figure}
\includegraphics[width=95mm, height=95mm]{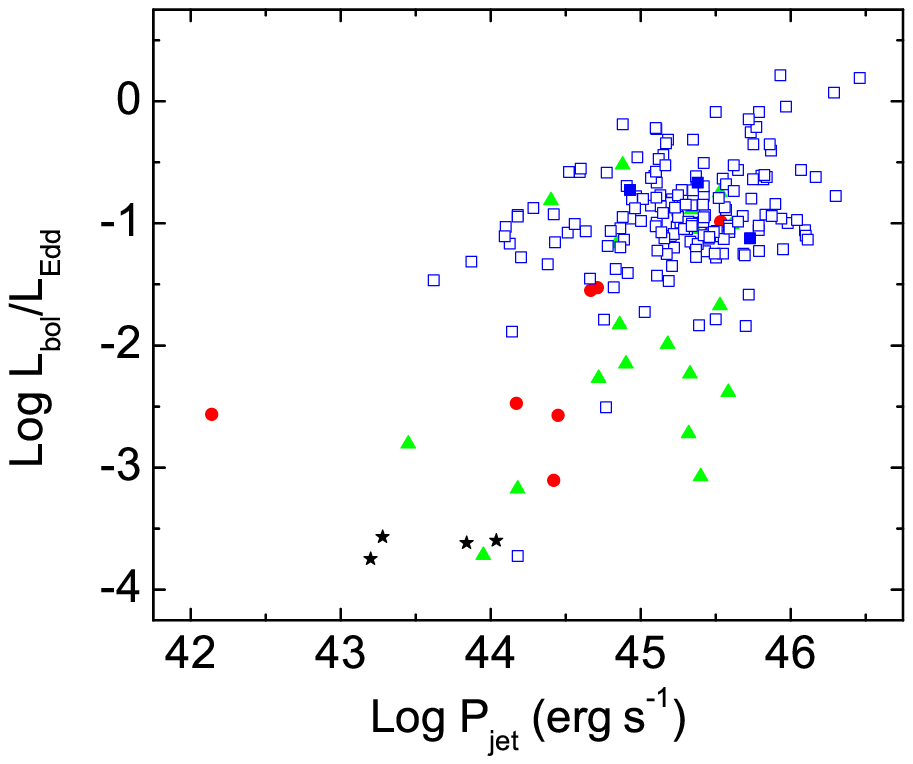}
\caption{The Eddington ratio as a function of jet power. The
uncertainty of jet kinetic power is 0.7 dex. The meanings of
different symbols are as same as Fig. 7.} \label{figure 10}
\end{figure}

\subsection{Jet power vs broad line luminosity and disk luminosity}

Figure 12 shows broad line luminosity as a function of jet power
(bottom panel) and disk luminosity as a function of jet power (top
panel, $L_{\rm d}\approx10L_{\rm BLR}$). Linear regression is
applied to the relevant data to analyze the correlation between
broad line luminosity and jet power. The results show a strong
correlation between broad line luminosity and jet power
($r=0.7\pm0.6, P<0.0001, N=226$). We also obtain $LogL_{\rm BLR}\sim
(0.98\pm0.07)LogP_{\rm jet}$. The result of Pearson partial analysis
shows that there is still significant correlation between broad line
luminosity and jet power ($N=217$, $P<0.0001$, $r=0.483$). From top
panel of Figure 12, we find that the distribution of data points is
close to $L_{\rm d}=P_{\rm jet}$. For almost all BL Lacs, the jet
power is larger than the disk luminosity while the jet power is much
smaller than the disk luminosity for most of FSRQs.

We use multiple linear regression analysis to get the relationship
between the jet power and both the Eddington luminosity and the
broad line region luminosity with 99\% confidence level and $r=0.71$
(Fig. 13):
\begin{equation}
LogP_{\rm jet}=0.52(\pm0.04)LogL_{\rm BLR}-0.02(\pm0.06)LogL_{\rm
Edd}+23.09(\pm2.4).
\end{equation}
After excluding these objects with black hole mass above $10^{9.5}
M_\odot$ and below $10^{7} M_\odot$, we get
\begin{equation}
LogP_{\rm jet}=0.51(\pm0.04)LogL_{\rm BLR}+0.02(\pm0.07)LogL_{\rm
Edd}+21.5(\pm2.8).
\end{equation}
Following Wang et al. (2004), we define the ``jet-dominance'' factor
(the relative importance of the jet power compared to the disk
luminosity) as $F_J=P_{\rm jet}/L_{\rm bol}$, Eddington ratio
$L_{\rm bol}/L_{\rm Edd}$ and $L_{\rm bol}\approx 10L_{BLR}$.
Equation (2) and (3) can be cast in a different form:
\begin{equation}
LogF_J=-0.5LogL_{\rm bol}+0.02LogL_{\rm bol}/L_{\rm Edd}+22.57.
\end{equation}
\begin{equation}
LogF_J=-0.47LogL_{\rm bol}-0.02LogL_{\rm bol}/L_{\rm
Edd}+20.99.
\end{equation}
This implies that ``jet-dominance'' is mainly controlled by, and is
inversely dependent on, the bolometric luminosity.

In addition, Equation (2) and (3) can be also expressed in a
different form as
\begin{equation}
LogP_{jet}=0.52LogL_{\rm bol}/L_{\rm Edd}+0.5Log(\frac{\rm M}{\rm
M_\odot})+41.62.
\end{equation}
\begin{equation}
LogP_{jet}=0.52LogL_{\rm bol}/L_{\rm Edd}+0.54Log(\frac{\rm M}{\rm
M_\odot})+43.15.
\end{equation}
Theoretically, Heinz \& Sunyaev (2003) have presented the dependence
of jet power on black hole mass and accretion rate in core dominated
jets: for standard accretion, $F_\nu\sim M^{17/12}$; for radiatively
inefficient accretion modes, $F_\nu\sim {(\dot{m}M)}^{17/12}$. The
observational evidence has been provided by many authors. There was
the black hole fundamental plane given by Merloni et al. (2003) and
Falcke et al. (2004): $Log L_R=(0.6^{+0.11}_{-0.11})Log
L_X+(0.78^{+0.11}_{-0.09})LogM+7.33^{+4.05}_{-4.07}$. Foschini
(2014) reported about the unification of relativistic jets from
compact objects. An important result from Foschini (2014) was the
discovery of powerful relativistic jets from radio-loud NLS1s, which
made it evident the existence of a secondary branch in AGN similar
to what was already known in Galactic binaries. From Foschini
(2014), in radiation-pressure dominated accretion disk, the jet
power can be scaled as $LogP_{\rm jet}\propto\frac{17}{12}LogM$; in
gas-pressure dominated accretion disk, $LogP_{\rm
jet}\propto\frac{17}{12}LogM+\frac{1}{2}Log\frac{L_{disk}}{L_{Edd}}$.
Through studying the jet power, radio loudness and black hole mass
in radio loud AGNs, Liu et al. (2006) found
$LogP_{jet}=0.22LogL_{\rm bol}/L_{\rm Edd}+0.59LogM+40.48$. Wang et
al. (2004) studied the properties of relativistic jets and obtained
$LogP_{\rm jet}=0.25(\pm0.09)LogL_{\rm BLR}+0.65(\pm0.25)LogL_{\rm
Edd}+5.07(\pm10.05)$. We compare our results with these results from
other authors, and find that our results are similar to results from
other authors, i.e., the dependence of jet power on both the
Eddington ratio and black hole mass. From Equations (6) and (7), we
can see that there are very close coefficients between black hole
mass and Eddington ratio. But from other results, the coefficient
from black hole mass is larger than that from Eddington ratio/X-ray
luminosity. This difference can be due to jet power calculated by
different methods and different sample.

\subsection{Jet power vs black hole mass and Eddington ratio}
We further analyze the correlations between jet power and black hole
mass, between jet power and Eddington ratio for all blazars (Fig.
14, 15). Similarly, excluding the redshift effect and using Pearson
partial analysis, we find that there are significant correlations
between jet power and black hole mass ($r=0.163, P=0.012, N=239$),
between jet power and Eddington ratio ($r=0.378, P<0.0001, N=208$).

\begin{figure}
\includegraphics[width=95mm, height=95mm]{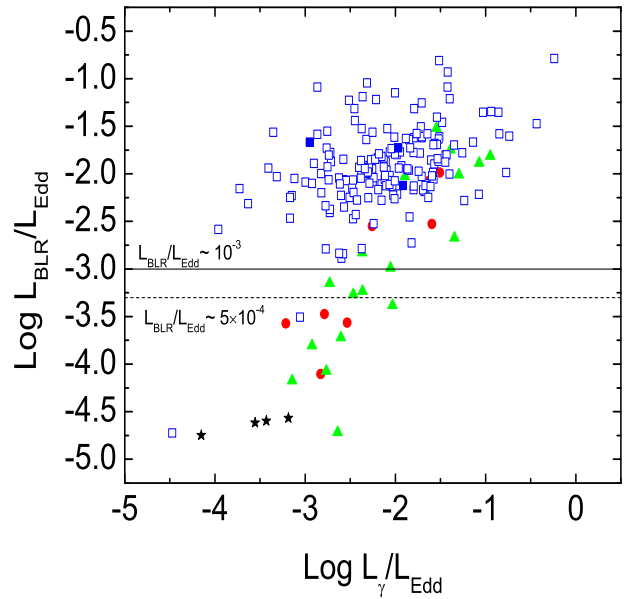}
\caption{Broad line luminosity as a function of $\gamma$-ray
luminosity both in Eddington units. The meanings of different
symbols are as same as Fig. 7. The horizontal solid line indicates
the luminosity divide between FSRQs and BL Lacs at $L_{\rm
BLR}/L_{\rm Edd}\sim10^{-3}$ and dashed line is $L_{\rm BLR}/L_{\rm
Edd}\sim5\times10^{-4}$ from Ghisellini et al. (2011) and Sbarrato
et al. (2012).} \label{figure 11}
\end{figure}

\begin{figure}
\includegraphics[width=95mm, height=95mm]{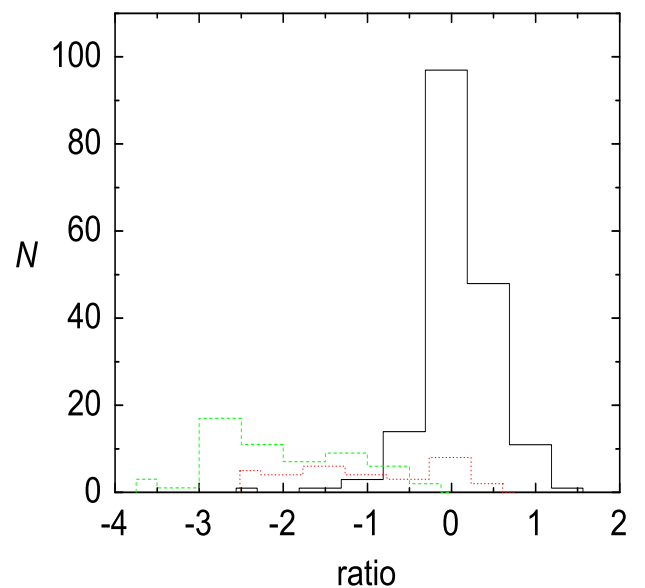}
\caption{Distributions of the accretion rates in Eddington units
$\dot{M}/\dot{M}_{\rm Edd}$. For FSRQs, $\dot{M}$ is given by
$\dot{M}=L_{\rm d}/(\eta c^2)$, with $\eta=0.1$ (black continuous
line) and for BL Lacs, $\dot{M}=P_{\rm jet}/c^2$ (green dashed line)
and $\dot{M}=L_{\rm d}/(\eta c^2)$ (red dotted line).} \label{figure
12}
\end{figure}

\begin{figure}
\includegraphics[width=95mm, height=200mm]{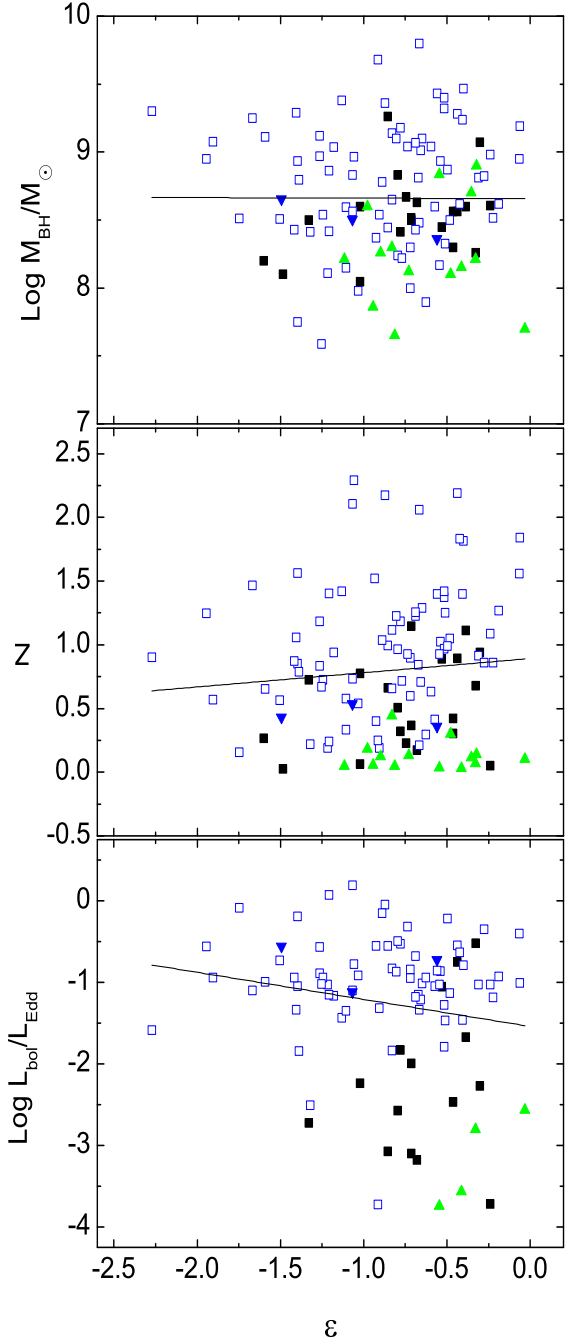}
\caption{Radiative efficiency of $\gamma$-ray
$\varepsilon=L_\gamma/(L_\gamma+P_{\rm jet})$ as a function of
redshift (middle panel), black hole mass (top panel) and Eddington
ratio (bottom panel). The black lines are the results of linear
regression. NFSRQs: blue empty squares; NBL Lacs: black filled
squares; TBL Lacs: green triangles in the positive direction;
TFSRQs: blue triangles in the negative direction. } \label{figure
16}
\end{figure}

\subsection{Divide between BL Lacs and FSRQs}

Ghisellini et al. (2011) and Sbarrato et al. (2012) have studied the
relation between $L_{\rm \gamma}/L_{\rm Edd}$ and $L_{\rm
BLR}/L_{\rm Edd}$, and proposed a physical distinction between the
two classes of blazars that division of blazars into BL Lacs and
FSRQs is controlled by the line luminosity in Eddington units. The
dividing line is of the order of $L_{\rm{BLR}}/L_{\rm{Edd}}\sim
5\times10^{-4}$, in good agreement with the idea that the presence
of strong emitting lines is related to a transition in the accretion
regime, becoming radiatively inefficient below a disk luminosity of
the order of one per cent of the Eddington one. With enlarged sample
and $\gamma$-ray excluding beaming effect, we revisit divide between
BL Lacs and FSRQs proposed by Ghisellini et al. (2011) and Sbarrato
et al. (2012). Fig. 16 is broad line luminosity as a function of
$\gamma$-ray luminosity both in Eddington units. Fig. 16 shows that
the divide between BL Lacs and FSRQs is of the order of $L_{\rm
BLR}/L_{\rm Edd}\sim10^{-3}$ corresponding to $\dot{M}/\dot{M}_{\rm
Edd}=0.1$ ($L_{\rm d}\approx10L_{\rm BLR}, \dot{M}_{\rm Edd}\equiv
L_{\rm Edd}/c^2, L_{\rm d}=\eta\dot{M}/c^2, \eta=0.1$). In Fig. 16,
we can see some transition sources with $L_{\rm BLR}/L_{\rm
Edd}>10^{-3}$ (transition sources are classified as BL Lacs with a
SED appearing as intermediate between BL Lacs and FSRQs, and also
have relatively weak broad emission lines and small EW). In
addition, Ghisellini et al. (2010) studied general physical
properties of bright Fermi blazars and found that there is a divide
between BL Lacs and FSRQs occurring at $\dot{M}/\dot{M}_{\rm
Edd}=0.1$. Following Ghisellini et al. (2010), we give the ratio
$\dot{M}/\dot{M}_{\rm
Edd}\equiv\frac{\dot{M}c^2}{1.3\times10^{38}{(M/M_\odot)}}$. For
FSRQs, $\dot{M}$ is given by $\dot{M}=L_{\rm d}/(\eta c^2)$, with
$\eta=0.1$ and for BL Lacs, $\dot{M}=P_{\rm jet}/c^2$. Our result is
given in Fig. 17 (black and green dashed lines). Meanwhile, we
consider $\dot{M}=L_{\rm d}/(\eta c^2)$ for BL Lacs. The result also
is given in Fig. 17 (black and red dotted lines). From Fig. 17, we
can see that the results are very similar to Fig. 16 that the divide
between BL Lacs and FSRQs occurs at $\dot{M}/\dot{M}_{\rm Edd}=0.1$
and some BL Lacs with $\dot{M}/\dot{M}_{\rm Edd}>0.1$ can be
transition sources. Therefore, the results from our sample (not
related to a particular model) confirm the idea proposed by
Ghisellini et al. (2010, 2011) and Sbarrato et al. (2012) and that
the divide between BL Lacs and FSRQs is of the order of $L_{\rm
BLR}/L_{\rm Edd}\sim10^{-3}$ corresponding to $\dot{M}/\dot{M}_{\rm
Edd}=0.1$.

\subsection{Radiative efficiency of $\gamma$-ray vs redshift, black hole mass and Eddington ratio}

The radiative efficiency of $\gamma$-ray is estimated as
$\varepsilon=L_\gamma/(L_\gamma+P_{\rm jet})$ (Nemmen et al. 2012).
When calculating radiative efficiency of $\gamma$-ray, we only use
jet power directly from Nemmen et al. (2012). Radiative efficiency
of $\gamma$-ray $\varepsilon=L_\gamma/(L_\gamma+P_{\rm jet})$ as a
function of redshift (middle panel), black hole mass (top panel) and
Eddington ratio (bottom panel) are shown in Fig. 18. Linear
regression is applied to the relevant data to analyze the
correlations ($N=110, r=0.09, P=0.37$; $N=109, r=-0.004, P=0.97$;
$N=90, r=-0.17, P=0.11$). From Fig. 18 and linear regression
analysis, we find that there are no correlations between radiative
efficiency of $\gamma$-ray and redshift, between radiative
efficiency of $\gamma$-ray and black hole mass, between radiative
efficiency of $\gamma$-ray and Eddington ratio. As we know,
according to theory of accretion disks, the accretion efficiency
depends on the radius of the innermost stable orbit, which in turn
depends on the spin of the black hole and the rotation of the
accretion disk. Also it is known that radiative efficiencies are
related with black hole spin.

\section{Discussions}

\subsection{The $\gamma$-ray luminosity}

The relation between $\gamma$-ray and broad line luminosity is
important for the origin of $\gamma$-ray. Sbarrato et al. (2012)
studied a Fermi sample and found a good correlation between the
luminosity of the broad lines and the $\gamma$-ray luminosity. But
they can not consider beaming effect for $\gamma$-ray luminosity and
the number of sample is still limited. Therefore, through
constructing a large sample of Fermi blazars and removing beaming
effect, we revisit the correlation and our results show that there
is significant correlation between intrinsic $\gamma$-ray and broad
line luminosity. Ghisellini \& Madau (1996) assessed non-thermal
Comptonization models for the high-energy emission of the EGRET
blazar sources, and found that the radiation produced by BLR clouds
illuminated by the relativistically moving plasma `blob' provides
the bulk of the seed photons to be Comptonization to $\gamma$-ray
energies. Through studying the connection between gamma-ray emission
and millimeter flares, Leon-Tavares et al. (2011b) found that the
mean observed delay from the beginning of a mm flare to the peak of
the $\gamma$-ray emission is about 70 days, corresponding to an
average distance of 7 parsecs along the jet. At these distances,
well beyond the canonical BLR, the seed photons could originate
either from the jet itself, from a dusty torus, or from an
outflowing BLR. Arshakian et al. (2010) suggested that the continuum
emission from the jet and counterjet ionizes material in a
subrelativistic outflow surrounding the jet, which results in a
formation of two conical regions with broad emission lines (in
addition to the conventional broad line region around the central
nucleus) at a distance $\geq0.4$ parsecs from the central engine.
The existence of a nonvirial, outflowing BLR can make EC models
possible even at distances of parsecs down the jet, which was first
proposed by Leon-Tavares et al. (2011b). Thus, the significant
correlation between intrinsic $\gamma$-ray and broad line luminosity
suggests that the radiation mechanism of the $\gamma$-ray in Fermi
blazars of existing BLR is likely to be inverse Compton scattering
of ambient photons from BLR or outflowing BLR. However, this result
can not totally exclude that the seed photons originate from other
sites. In addition, we also find significant correlations between
intrinsic $\gamma$-ray luminosity and black hole mass, between
intrinsic $\gamma$-ray luminosity and Eddington ratio, which are
consistent with the results of jet power. According to relativistic
jet theory, the radiative jet power can be calculated by dividing
the observed gamma-ray luminosity by the square of the bulk Lorentz
factor. Therefore, it is known that $\gamma$-ray luminosity can be
used as a proxy for the jet power.

\subsection{Jet power}

From our results, we can see that the correlation between broad line
luminosity and jet power is significant which supports that jet
power has a close link with accretion. According to Ghisellini
(2006), if relativistic jets are powered by a Poynting flux, under
some reasonable assumption, the BZ jet power can be written
\begin{equation}
L_{\rm{BZ,jet}}\sim(\frac{a}{m})^{2}\frac{R_{\rm
S}^{3}}{HR^{2}}\frac{\varepsilon_{\rm B}}{\eta}\frac{L_{\rm
disk}}{\beta_{\rm r}},
\end{equation}
where $\frac{a}{m}$ is the specific BH angular momentum; $R_{\rm
S}=\frac{2GM_{\rm BH}}{C^{2}}$ is the Schwarzschild radius; $H$ is
the disk thickness; $R$ is the radius; $\varepsilon_{\rm B}$ is the
fraction of the available gravitational energy; $\eta$ is the
accretion efficiency; $L_{\rm disk}$ is the observed luminosity of
accretion disk; $\beta_{\rm r}$ is the radial infalling velocity.
The maximum BZ jet power can then be written as (Ghisellini 2006)
\begin{equation}
L_{\rm jet}\sim\frac{L_{\rm disk}}{\eta}.
\end{equation}
In addition, in view of current theories of accretion disks, the BLR
is ionized by the radiation of the accretion disk. We have
\begin{equation}
L_{\rm disk}\approx10L_{\rm BLR}.
\end{equation}
From equation (9) and (10), we have
\begin{equation}
 L_{\rm BLR}\sim0.1\eta L_{\rm jet}.
\end{equation}
From equation (11), we can have
\begin{equation}
 LogL_{\rm BLR}=LogL_{\rm jet}+Log\eta+const.
\end{equation}
From equation (12), we can find that the theoretical predicted
coefficient of $LogL_{\rm BLR}-LogL_{\rm jet}$ relation is 1. Using
linear regression analysis, we obtain $LogL_{\rm
BLR}\sim(0.98\pm0.07)Log P_{\rm jet}$ for all blazars, which is
consistent with the theoretical predicted coefficient of $LogL_{\rm
BLR}-LogL_{\rm jet}$ relation. Then our results suggest that Fermi
blazars jets are also powered by energy extraction from a rapidly
spinning black hole through the magnetic field provided by the
accretion disk, which supports the hypothesis provided by Xie et al.
(2006, 2007). The extraction of energy from black hole rotation was
well established by Blandford \& Znajek (1977). In addition, we find
that the jet power depends on both the Eddington ratio and black
hole mass. Heinz \& Sunyaev (2003) have presented the theoretical
dependence of jet power on Eddington ratio and black hole mass. The
observational evidence has been provided by many authors (see
Section 3.4). The massive black holes will be spun up through
accretion, as the black holes acquire mass and angular momentum
simultaneously through accretion (Chai et al. 2012). Volonteri et
al. (2007) investigated how the accretion from a warped disc
influences the evolution of black hole spins with the effects of
accretion and merger being properly considered and concluded that
within the cosmological framework, most supermassive black holes in
elliptical galaxies have on average higher spins than black holes in
spiral galaxies, where random, small accretion episodes (e.g.,
tidally disrupted stars, accretion of molecular clouds) might have
played a more important role. If this is true, the correlation
between black hole mass and jet power implies that jet power is
probably governed by the black hole spin. So from above discussion,
we can conclude that for Fermi blazars, jets are powered by energy
extraction from both accretion and black hole spin (i.e., not by
accretion only).

From top panel of Figure 12, we find that for almost all BL Lacs,
the jet power is larger than the disk luminosity while the jet power
is much smaller than the disk luminosity for most of FSRQs. For BL
Lacs, our result is consistent with result of Ghisellini et al.
(2010), whereas for FSRQs, our result is different from result of
Ghisellini et al. (2010) in which jet power is still larger than the
disk luminosity. In their work, the jet power and disk luminosity
are related with the model described in detail in Ghisellini \&
Tavecchio (2009c). However, our results are model-independent and
much larger sample. A reasonable explanation about our results is as
follows. FSRQs occur in the earlier phase. They have powerful disk
and jet, high accretion and $L_{\rm d}>P_{\rm jet}$. With time, the
FSRQs will have lower accretion rate, a less efficient disk,
shrinking BLR. It is possible that some transitions between FSRQs
and BL Lacs appear with moderate BLR luminosity. When the accretion
rate decreases below the critical value (i.e., $\dot{m}_{\rm
c}=\dot{M}_{\rm div}/\dot{M}_{\rm Edd}\sim 10^{-1}$), the accretion
changes mode, becoming radiatively inefficient and that FSRQs become
BL Lacs. BL Lacs have weak disk and weaker lines emitted closer to
the black hole. Dissipation in the jet occurs outside the BLR (if it
exists at all). So it is possible that BL Lacs have $L_{\rm
d}<P_{\rm jet}$ and the explanation is in line with scenario
proposed by Cavaliere \& D'Elia (2002) and Ghisellini \& Tavecchio
(2008). Another result of this study is that ``jet-dominance'' is
mainly controlled by, and is inversely dependent on, the bolometric
luminosity from equation (4) and (5). Wang et al. (2004) studied a
sample of 35 blazars, and found that the ``jet-dominance'' is mainly
controlled by, and is inversely dependent on, the Eddington ratio.
In our study, our sample is much larger than that of Wang et al.
(2004) and only focuses on Fermi sample; jet power is estimated by
different method. These can explain the different results between
us.

\subsection{TeV blazars}

In this subsection, we discuss the properties of TeV blazars
detected by Fermi LAT. Fig. 1-6 are distributions of redshift, black
hole mass, jet kinetic power, intrinsic $\gamma$-ray luminosity,
$\gamma$-ray photon index and bulk Lorentz factor. We find that
compared with NBL Lacs, TBL Lacs have much smaller redshift, jet
kinetic power, intrinsic $\gamma$-ray luminosity, $\gamma$-ray
photon index and bulk Lorentz factor. And the distributions of these
parameters between them are significant difference. However, there
are not significant differences of black hole masses between them.
Due to most of TBL Lacs classed into HBLs, we also compare the
distributions of these parameters among HBLs, IBLs, LBLs. The
results show that except black hole mass distributions, there are
significant differences for these parameters distributions between
HBLs and IBLs, between HBLs and LBLs. The TBL Lacs are relatively
nearby blazars ($z<0.5$, $z\approx0.1$ for most of them) because of
TeV $\gamma$-ray absorbed by extragalactic background light. No
significant differences of black hole masses between TBL Lacs and
NBL Lacs suggest that black hole mass is not main factor for
difference between them. For TBL Lacs, $\gamma$-ray photon index
$\Gamma_{\rm GeV}<2.2$, which is consistent with the results of
Senturk et al. (2013). Compared with NBL Lacs, TBL Lacs have much
smaller jet kinetic power and intrinsic $\gamma$-ray luminosity
which suggests that TBL Lacs mainly is low power sources and there
are different jet structures between TBL Lacs and NBL Lacs. In our
sample, the mean radio bulk Lorentz factor of TBL Lacs is
6.27$\pm$0.58. Compared with NBL Lacs, TBL Lacs have much smaller
bulk Lorentz factor. Many authors have studied the parsec-scale jets
of the TBL Lacs and found a lower Doppler factor, bulk Lorentz
factor and slower apparent jet pattern speeds (e.g., Piner et al.
2008, 2010, 2013; Kovalev et al. 2005; Giroletti et al. 2004a;
Chiaberge et al. 2000). For TBL Lacs, there is `bulk Lorentz factor
crisis'. Doppler factors from SSC models are in strong disagreement
with those deduced from the unification models between blazars and
radio galaxies. When corrected from extragalactic absorption by the
diffuse infrared background, the SSC one-zone models require very
high Lorentz factor (around 50) to avoid strong $\gamma-\gamma$
absorption. However, the statistics on beamed vs unbeamed objects,
as well as the luminosity contrast, favor much lower Lorentz factor
of the order of 3 (Henri \& Sauge 2006). An obvious explanation for
the `bulk Lorentz factor crisis' is that the radio and gamma-ray
emissions are produced in different parts of the jet with different
bulk Lorentz factors and several models have been invoked, including
decelerated jets, spine-sheath structures, faster moving leading
edges of blobs, and `minijets' within the main jet (Piner et al.
2013). All models suggest that jet of TBL Lacs have significant
velocity structures. The velocity structures may show an
observational signature in the VLBI image of jet, such as
limb-brightening or limb-darkening. Limb-brightening has been
observed in VLBI images of Mrn 501 and Mrn 421 (e.g., Giroletti et
al. 2004b; Piner \& Edward 2005). Therefore, our statistical results
strengthen that TeV BL Lacs have a low bulk Lorentz factor for the
parsec-scale radio emission. Finally, For TFSRQs, we find that
compared with normal GeV FSRQs (NFSRQs), TFSRQS have much smaller
redshift but larger bulk Lorentz factor which suggests that the jets
of TFSRQs have a high bulk Lorentz factor for parsec-scale radio
emission.

\section{Conclusions}

In this work we have analyzed a large sample of blazars detected in
the Fermi satellite. Our main results are the following:

(i) After excluding beaming effect and redshift effect, there is
significant correlation between intrinsic $\gamma$-ray and broad
line luminosity which suggests that the radiation mechanism of the
$\gamma$-ray in Fermi blazars of existing BLR is likely to be
inverse Compton scattering of ambient photons from BLR or outflowing
BLR. And there are significant correlations between intrinsic
$\gamma$-ray luminosity and black hole mass, between intrinsic
$\gamma$-ray luminosity and Eddington ratio.

(ii) The results from our sample (not related to a particular model)
confirm the idea proposed by Ghisellini et al. (2010, 2011) and
Sbarrato et al. (2012) and that the divide between BL Lacs and FSRQs
is of the order of $L_{\rm BLR}/L_{\rm Edd}\sim10^{-3}$
corresponding to $\dot{M}/\dot{M}_{\rm Edd}=0.1$.

(iii) The correlation between broad line luminosity and jet power is
significant which supports that jet power has a close link with
accretion. Jet power depends on both the Eddington ratio and black
hole mass. We also obtain $LogL_{\rm BLR}\sim(0.98\pm0.07)Log P_{\rm
jet}$ for all blazars, which is consistent with the theoretical
predicted coefficient of $LogL_{\rm BLR}-LogL_{\rm jet}$ relation.
These results support that for Fermi blazar, jets are powered by
energy extraction from both accretion and black hole spin (i.e., not
by accretion only).

(iv) For almost all BL Lacs, the jet power is larger than the disk
luminosity while the jet power is much smaller than the disk
luminosity for most of FSRQs. The ``jet-dominance'' is mainly
controlled by, and is inversely dependent on, the bolometric
luminosity.

(v) There are no correlations between radiative efficiency of
$\gamma$-ray and redshift, between radiative efficiency of
$\gamma$-ray and black hole mass, between radiative efficiency of
$\gamma$-ray and Eddington ratio.

(vi) Compared with NBL Lacs, TBL Lacs have much smaller redshift,
jet kinetic power, intrinsic $\gamma$-ray luminosity, $\gamma$-ray
photon index and bulk Lorentz factor for parsec-scale radio
emission. There are not significant differences of black hole masses
between them. TFSRQS have small redshift but large bulk Lorentz
factor for parsec-scale radio emission.

\section*{Acknowledgments}

We sincerely thank anonymous referee for valuable comments and
suggestions. We also thank Minfeng Gu for helpful suggestions. This
work is financially supported by the National Nature Science
Foundation of China (11163007, U1231203). This research has made use
of the NASA/IPAC Extragalactic Database (NED), that is operated by
Jet Propulsion Laboratory, California Institute of Technology, under
contract with the National Aeronautics and Space Administration.

\begin{table*}
 \centering
  \caption{The sample.}
\begin{tabular}{@{}llrrrrlrlr@{}}
  \hline\hline
   Fermi name   &   Other name  &   RA  &   Opt.type    &   Redshift    &   $LogL_{\rm \gamma}^{int}$   &   $LogM_{\rm BH}$    &   $LogP_{\rm jet}$   &   $LogL_{\rm BLR}$   \\
    &       &   DEC &   SED type    &      &   Unc. &   ref & $Logf_{\rm b}$   &   ref \\
(1) &   (2) &   (3) &   (4) &   (5) &   (6) &   (7) &   (8) &   (9) \\
 \hline
2FGL J0004.7-4736   &   PKS 0002-478    &   00 04 35.5  &   bzq &   0.88    &   44.36   &   7.85    &   45.07   &   44.11   \\
    &       &   -47 36 20.4 &   LSP &       &   0.57    &   Sh12    &   -2.44   &   Sh12    \\
2FGL J0017.4-0018   &   S3 0013-00  &   00 16 10.9  &   bzq &   1.574   &   44.64   &   8.55,9.04   &   45.35   &   44.91   \\
    &       &   -00 15 14.5 &   LSP &       &   0.57    &   S11 &   -2.63   &   S11 \\
2FGL J0017.6-0510   &   PMN 0017-0512   &   00 17 35.6  &   bzq &   0.227   &   43.56   &   7.55    &   44.29   &   43.79   \\
\hline
\end{tabular}
\end{table*}

\addtocounter{table}{-1}
\begin{table*}
 \centering
  \caption{$Continued.$}
  \begin{tabular}{@{}llrrrrlrlr@{}}
  \hline\hline
  Fermi name   &   Other name  &   RA  &   Opt.type    &   Redshift    &   $LogL_{\rm \gamma}^{int}$   &   $LogM_{\rm BH}$    &   $LogP_{\rm jet}$   &   $LogL_{\rm BLR}$   \\
    &       &   DEC &   SED type    &      &   Unc. &   ref & $Logf_{\rm b}$   &   ref \\
(1) &   (2) &   (3) &   (4) &   (5) &   (6) &   (7) &   (8) &   (9) \\
 \hline
2FGL J0319.6+1849   &   0317+185    &     03 19 51.8    &   tbzb    &   0.19    &   43.37   &   8.10    &   44.11   &       \\
    &       &    +18 45 34  &   HSP &       &   0.57    &   X04 &   -1.81   &       \\
2FGL J0326.1+2226   &   TXS 0322+222    &   03 25 36.7  &   bzq &   2.066   &   45.12   &   9.5,9.16    &   45.82   &   45.81   \\
    &       &   +22 24 01.0 &   LSP &       &   0.57    &   Sh12    &   -2.93   &   Sh12    \\
2FGL J0339.4-0144   &   PKS 0336-01 &     03 39 30.9    &   bzq &   0.852   &   43.93   &   8.89,8.98   &   45.31$^\dagger$   &   45.00   \\
\hline
\end{tabular}
\end{table*}

\addtocounter{table}{-1}
\begin{table*}
 \centering
  \caption{$Continued.$}
  \begin{tabular}{@{}llrrrrlrlr@{}}
  \hline\hline
  Fermi name   &   Other name  &   RA  &   Opt.type    &   Redshift    &   $LogL_{\rm \gamma}^{int}$   &   $LogM_{\rm BH}$    &   $LogP_{\rm jet}$   &   $LogL_{\rm BLR}$   \\
    &       &   DEC &   SED type    &      &   Unc. &   ref & $Logf_{\rm b}$   &   ref \\
(1) &   (2) &   (3) &   (4) &   (5) &   (6) &   (7) &   (8) &   (9) \\
 \hline
2FGL J0635.5-7516   &   PKS 0637-75 &     06 35 46.5    &   bzq &   0.653   &   44.40   &   9.41,8.81   &   45.98$^\dagger$   &   45.23   \\
    &       &    -75 16 17  &       &       &   0.56    &   W02,L06 &   -2.49   &   C99 \\
2FGL J0654.2+4514   &   B3 0650+453 &   06 54 23.6  &   bzq &   0.928   &   44.70   &   8.17    &   45.10$^\dagger$   &   44.26   \\
    &       &   +45 14 23.2 &   LSP &       &   0.56    &   Sh12    &   -2.65   &   Sh12    \\
2FGL J0654.5+5043   &   GB6 J0654+5042  &   06 54 22.0  &   bzq &   1.253   &   43.27   &   7.86,8.79   &   43.62$^\dagger$   &   43.97   \\
\hline
\end{tabular}
\end{table*}

\addtocounter{table}{-1}
\begin{table*}
 \centering
  \caption{$Continued.$}
  \begin{tabular}{@{}llrrrrlrlr@{}}
  \hline\hline
  Fermi name   &   Other name  &   RA  &   Opt.type    &   Redshift    &   $LogL_{\rm \gamma}^{int}$   &   $LogM_{\rm BH}$    &   $LogP_{\rm jet}$   &   $LogL_{\rm BLR}$   \\
    &       &   DEC &   SED type    &      &   Unc. &   ref & $Logf_{\rm b}$   &   ref \\
(1) &   (2) &   (3) &   (4) &   (5) &   (6) &   (7) &   (8) &   (9) \\
 \hline
2FGL J0920.9+4441   &   B3 0917+449 &    09 20 58.4 &   bzq &   2.19    &   45.60   &   9.25,9.31,9.29  &   45.84$^\dagger$   &   45.85,45.7  \\
    &       &    +44 41 54  &   LSP &       &   0.56    &   Sb12,C12,Sh12   &   -3.12   &   Sb12,Sh12   \\
2FGL J0921.9+6216   &   OK 630  &   09 21 36.1  &   bzq &   1.453   &   44.71   &   8.93    &   45.42   &   45.05   \\
    &       &   +62 15 51.1 &   LSP &       &   0.57    &   Sh12    &   -2.67   &   Sh12    \\
2FGL J0923.2+4125   &   B3 0920+416 &   09 23 31.1  &   bzq &   1.732   &   44.80   &   7.68,8.16   &   45.50   &   43.75   \\
\hline
\end{tabular}
\end{table*}

\addtocounter{table}{-1}
\begin{table*}
 \centering
  \caption{$Continued.$}
  \begin{tabular}{@{}llrrrrlrlr@{}}
  \hline\hline
  Fermi name   &   Other name  &   RA  &   Opt.type    &   Redshift    &   $LogL_{\rm \gamma}^{int}$   &   $LogM_{\rm BH}$    &   $LogP_{\rm jet}$   &   $LogL_{\rm BLR}$   \\
    &       &   DEC &   SED type    &      &   Unc. &   ref & $Logf_{\rm b}$   &   ref \\
(1) &   (2) &   (3) &   (4) &   (5) &   (6) &   (7) &   (8) &   (9) \\
 \hline
2FGL J1130.3-1448   &   1127-145    &     11 30 07.0    &   bzq &   1.184   &   44.92   &   9.18    &   45.62$^\dagger$   &   45.77   \\
    &       &    -14 49 27  &   LSP &       &   0.57    &   C12 &   -2.76   &   Z09 \\
2FGL J1136.7+7009   &   Mrk 180 &     11 36 26.4    &   tbzb    &   0.045278    &   42.40   &   8.21    &   43.48$^\dagger$   &       \\
    &       &    +70 09 27  &   HSP &       &   0.56    &   W02 &   -1.45   &       \\
2FGL J1146.8-3812   &   PKS 1144-379    &     11 47 01.4    &   bzq &   1.048   &   44.58   &   8.50    &   44.89$^\dagger$   &   44.36,44.60 \\
\hline
\end{tabular}
\end{table*}

\addtocounter{table}{-1}
\begin{table*}
 \centering
  \caption{$Continued.$}
  \begin{tabular}{@{}llrrrrlrlr@{}}
  \hline\hline
  Fermi name   &   Other name  &   RA  &   Opt.type    &   Redshift    &   $LogL_{\rm \gamma}^{int}$   &   $LogM_{\rm BH}$    &   $LogP_{\rm jet}$   &   $LogL_{\rm BLR}$   \\
    &       &   DEC &   SED type    &      &   Unc. &   ref & $Logf_{\rm b}$   &   ref \\
(1) &   (2) &   (3) &   (4) &   (5) &   (6) &   (7) &   (8) &   (9) \\
 \hline
2FGL J1332.5-1255   &   PMN J1332-1256  &   13 32 39.1  &   bzq &   1.492   &   45.05   &   8.96,8.61   &   45.75   &   45.26   \\
    &       &   -12 56 15.7 &   LSP &       &   0.57    &   Sh12    &   -2.89   &   Sh12    \\
2FGL J1332.7+4725   &   B3 1330+476 &    13 32 45.2 &   bzq &   0.668   &   43.91   &   8.56,7.975  &   44.63   &   44.41,44.22 \\
    &       &    +47 22 23  &   LSP &       &   0.57    &   Sb12,Sh12   &   -2.16   &   Sb12    \\
2FGL J1333.5+5058   &   CLASS J1333+5057    &   13 33 53.8  &   bzq &   1.362   &   44.71   &   7.95    &   45.42   &   44.37   \\
\hline
\end{tabular}
\end{table*}

\addtocounter{table}{-1}
\begin{table*}
 \centering
  \caption{$Continued.$}
  \begin{tabular}{@{}llrrrrlrlr@{}}
  \hline\hline
  Fermi name   &   Other name  &   RA  &   Opt.type    &   Redshift    &   $LogL_{\rm \gamma}^{int}$   &   $LogM_{\rm BH}$    &   $LogP_{\rm jet}$   &   $LogL_{\rm BLR}$   \\
    &       &   DEC &   SED type    &      &   Unc. &   ref & $Logf_{\rm b}$   &   ref \\
(1) &   (2) &   (3) &   (4) &   (5) &   (6) &   (7) &   (8) &   (9) \\
 \hline
2FGL J1550.7+0526   &   4C+05.64    &     15 50 35.3    &   bzq &   1.422   &   44.69   &   9.38,8.98   &   45.79$^\dagger$   &   45.06,45.08 \\
    &       &    +05 27 10  &   LSP &       &   0.57    &   Sb12,Sh12   &   -2.64   &   Sb12,Sh12   \\
2FGL J1553.5+1255   &   PKS 1551+130    &     15 53 32.7    &   bzq &   1.29    &   44.98   &   9.1,8.64    &   45.52$^\dagger$   &   45.20,45.18 \\
    &       &    +12 56 52  &   LSP &       &   0.57    &   Sb12,Sh12   &   -2.8    &   Sb12,Sh12   \\
2FGL J1608.5+1029   &   4C+10.45    &     16 08 46.2    &   bzq &   1.226   &   44.70   &   8.64,9.5,8.77   &   45.29$^\dagger$   &   45.01,45.07 \\
\hline
\end{tabular}
\end{table*}

\addtocounter{table}{-1}
\begin{table*}
 \centering
  \caption{$Continued.$}
  \begin{tabular}{@{}llrrrrlrlr@{}}
  \hline\hline
  Fermi name   &   Other name  &   RA  &   Opt.type    &   Redshift    &   $LogL_{\rm \gamma}^{int}$   &   $LogM_{\rm BH}$    &   $LogP_{\rm jet}$   &   $LogL_{\rm BLR}$   \\
    &       &   DEC &   SED type    &      &   Unc. &   ref & $Logf_{\rm b}$   &   ref \\
(1) &   (2) &   (3) &   (4) &   (5) &   (6) &   (7) &   (8) &   (9) \\
 \hline
2FGL J2000.0+6509   &   1ES 1959+650    &    19 59 59.8 &   tbzb    &   0.047   &   42.97   &   8.10    &   43.71   &       \\
    &       &    +65 08 55  &   HSP &       &   0.57    &   Z12 &   -1.55   &       \\
2FGL J2000.8-1751   &   PKS 1958-179    &     20 00 57.1    &   bzq &   0.65    &   44.37   &       &   45.08   &   44.15   \\
    &       &    -17 48 58  &   LSP &       &   0.57    &       &   -2.45   &   C99 \\
2FGL J2009.5-4850   &   PKS 2005-489    &     20 09 25.4    &   tbzb    &   0.071   &   43.10   &   8.5,9.03,8.1    &   43.84   &   42.04   \\
\hline
\end{tabular}
\end{table*}

\addtocounter{table}{-1}
\begin{table*}
 \centering
  \caption{$Continued.$}
  \begin{tabular}{@{}llrrrrlrlr@{}}
  \hline\hline
  Fermi name   &   Other name  &   RA  &   Opt.type    &   Redshift    &   $LogL_{\rm \gamma}^{int}$   &   $LogM_{\rm BH}$    &   $LogP_{\rm jet}$   &   $LogL_{\rm BLR}$   \\
    &       &   DEC &   SED type    &      &   Unc. &   ref & $Logf_{\rm b}$   &   ref \\
(1) &   (2) &   (3) &   (4) &   (5) &   (6) &   (7) &   (8) &   (9) \\
 \hline
2FGL J2315.7-5014   &   PKS 2312-505    &   23 15 44.2  &   bzb &   0.808   &   44.12   &   7.68    &   44.84   &   43.63   \\
    &       &   -50 18 39.7 &   LSP &       &   0.57    &   Sh12    &   -2.29   &   Sh12    \\
2FGL J2322.2+3206   &   B2 2319+31  &   23 21 54.7  &   bzq &   1.489   &   44.75   &   8.66,8.75   &   45.46   &   44.71   \\
    &       &   +32 04 05.5 &   LSP &       &   0.57    &   Sh12    &   -2.70   &   Sh12    \\
2FGL J2327.5+0940   &   PKS 2325+093    &   23 27 33.4  &   bzq &   1.841   &   45.18   &   8.7,9.35    &   45.87   &   45.20   \\
\hline
\end{tabular}
\begin{quote}
(a) References. C99: Cao \& Jiang (1999); C12: Chai et al. (2012);
G01: Gu et al. (2001); L06: Liu et al.
  (2006); L11: Leon-Tavares et al. (2011a); Sb12: Sbarrato et al. (2012); Sh12: Shaw et al. (2012); S11: Shen
et al. (2011); W04: Wang et al. (2004); W02: Woo \& Urry (2002);
X91, X04: Xie et al. (1991, 2004); Z12: Zhang et al. (2012); Z09:
Zhou \& Cao (2009).

(b) Values with a `` $\ast$ '' correspond to $f_b$ measured by the
radio measurements, otherwise they were estimated from the best-fit
of Nemmen et al. (2012).

(c) Values with a `` $\dagger$ '' correspond to $P_{\rm jet}$
estimated by using the correlation between the extended radio
emission and the jet power, otherwise they were estimated from the
best-fit of Nemmen et al. (2012).
\end{quote}
\end{table*}


\begin{thebibliography}{99}
\bibitem[\protect\citeauthoryear{Abdo et al.}{2009}]{b1} Abdo A.A., Ackermann M., Ajello M. et al., 2009, ApJ, 700, 597
\bibitem[\protect\citeauthoryear{Abdo et al.}{2010a}]{b2} Abdo A.A., Ackermann M., Ajello M. et al., 2010a, ApJS, 188, 405
\bibitem[\protect\citeauthoryear{Abdo et al.}{2010b}]{b3} Abdo A.A., Ackermann M., Ajello M. et al., 2010b, ApJ, 715, 429
\bibitem[\protect\citeauthoryear{Abdo et al.}{2010c}]{b4} Abdo A.A., Ackermann M., Agudo I. et al., 2010c, ApJ, 716, 30
\bibitem[\protect\citeauthoryear{Abdo et al.}{2012}]{b6} Abdo A.A., Ackermann M., Ajello M. et al., 2012, ApJS, 199, 31
\bibitem[\protect\citeauthoryear{Ackermann et al.}{2011a}]{b7} Ackermann M., Ajello M., Allafort A. et al., 2011a, ApJ, 743, 171
\bibitem[\protect\citeauthoryear{Ackermann et al.}{2011b}]{b8} Ackermann M., Ajello M., Allafort A. et al., 2011b, ApJ, 741, 30
\bibitem[\protect\citeauthoryear{Aharonian et al.}{2007}]{b8} Aharonian F. et al., 2007, ApJ, 664, L71
\bibitem[\protect\citeauthoryear{Aharonian et al.}{2007}]{b8} Albert J. et al., 2006, ApJ, 639, 761
\bibitem[\protect\citeauthoryear{Allen et al.}{2006}]{b9} Allen S.W., Dunn R.J.H., Fabian A.C., Taylor G.B. \& Reynolds C.S., 2006, MNRAS, 372, 21
\bibitem[\protect\citeauthoryear{Allen et al.}{2006}]{b9} Arshakian T.G. et al., 2010, MNRAS, 401, 1231
\bibitem[\protect\citeauthoryear{Balmaverde et al.}{2008}]{b10} Balmaverde B., Baldi R.D. \& Capetti A., 2008, A\&A, 486, 119
\bibitem[\protect\citeauthoryear{Balmaverde et al.}{2008}]{b10} Blandford R.D., \& Znajek R.L., 1977, MNRAS, 179, 433
\bibitem[\protect\citeauthoryear{Blandford \& Rees}{1978}]{b12} Blandford R.D., Rees M.J., 1978, in Pittsburgh Conference on BL Lac Objects. Pittsburgh Univ., PA, p.328
\bibitem[\protect\citeauthoryear{Balmaverde et al.}{2008}]{b10} Blandford R.D., \& Payne D.G., 1982, MNRAS, 199, 883
\bibitem[\protect\citeauthoryear{Cavagnolo}{2010}]{b13} Cavagnolo K.W. et al., 2010, ApJ, 720, 1066
\bibitem[\protect\citeauthoryear{Cavaliere \& D'Elia}{2002}]{b13} Cavaliere A., \& D'Elia V., 2002, ApJ, 571, 226
\bibitem[\protect\citeauthoryear{Cao \& Jiang}{1999}]{b14} Cao X., \& Jiang D.R., 1999, MNRAS, 307, 802
\bibitem[\protect\citeauthoryear{Celotti \& Fabian}{1993}]{b15} Celotti A., \& Fabian A.C., 1993, MNRAS, 264, 228
\bibitem[\protect\citeauthoryear{Celotti et al.}{1997}]{b16} Celotti A., Padovani P., \& Ghisellini G., 1997, MNRAS, 286, 415
\bibitem[\protect\citeauthoryear{Celotti et al.}{2001}]{b17} Celotti A., Ghisellini G., \& Chiaberge M., 2001, MNRAS, 321, L1
\bibitem[\protect\citeauthoryear{Chai et al.}{2012}]{b19} Chai B., Cao X., \& Gu M., 2012, ApJ, 759, 114
\bibitem[\protect\citeauthoryear{Cheng et al.}{1993}]{b20} Cheng K.S., Yu K.N., \& Ding K.Y., 1993, A\&A, 275, 53
\bibitem[\protect\citeauthoryear{Cheng \& Ding}{1994}]{b21} Cheng K.S., \& Ding K.Y., 1994, A\&A, 288, 97
\bibitem[\protect\citeauthoryear{Cheng \& Ding}{1994}]{b21} Chiaberge M., Celotti A., Capetti A., \& Ghisellini G., 2000, A\&A, 358, 104
\bibitem[\protect\citeauthoryear{Cheng \& Ding}{1994}]{b21} Costamante L., \& Ghisellini G., 2002, A\&A, 384, 56
\bibitem[\protect\citeauthoryear{De Angelis et al.}{2008}]{b21} De Angelis A., Mansutti O., Persic M., 2008 La Rivista del Nuovo Cimento, 31, 187 (arXiv:1011.6660)
\bibitem[\protect\citeauthoryear{Dermer et al.}{1992}]{b22} Dermer C.D., Schlickeiser R., \& Mastichiadis A., 1992, A\&A, 256, L27
\bibitem[\protect\citeauthoryear{Falcke \& Biermann}{1995}]{b23} Falcke H., \& Biermann P.L., 1995, A\&A, 293, 665
\bibitem[\protect\citeauthoryear{Falcke \& Biermann}{1995}]{b23} Falcke H. et al., 2004, A\&A, 414, 895
\bibitem[\protect\citeauthoryear{Fichtel et al.}{1994}]{b24} Fichtel C.E., Bertsch D.L., Chiang J. et al., 1994, ApJS, 94, 551
\bibitem[\protect\citeauthoryear{Fichtel et al.}{1994}]{b24} Foschini L., 2014, IJMPS, 2860188F.
\bibitem[\protect\citeauthoryear{Francis et al.}{1991}]{b26} Francis P.J., Hewett P.C., Foltz C.B., Chaffee F.H., Weymann R.J. \& Morris S.L., 1991, ApJ, 373, 465
\bibitem[\protect\citeauthoryear{Ghisellini et al.}{1993}]{b27} Ghisellini G. et al., 1993, ApJ, 407, 65
\bibitem[\protect\citeauthoryear{Ghisellini \& Madau}{1996}]{b28} Ghisellini G., \& Madau P., 1996, MNRAS, 280, 67
\bibitem[\protect\citeauthoryear{Ghisellini}{2006}]{b31} Ghisellini G., 2006, in Proc. VI Microquasar Workshop: Microquasars and Beyond, http://pos.sissa.it//archive/conferences/033/027/MQW6\_027.pdf
\bibitem[\protect\citeauthoryear{Ghisellini \& Tavecchio}{2008}]{b32} Ghisellini G., \& Tavecchio F., 2008, MNRAS, 387, 1669
\bibitem[\protect\citeauthoryear{Ghisellini et al.}{2009a}]{b33} Ghisellini G., Tavecchio F., \& Ghirlanda G., 2009a, MNRAS, 399, 2041
\bibitem[\protect\citeauthoryear{Ghisellini et al.}{2009b}]{b34} Ghisellini G., Tavecchio F., Foschini L., Ghirlanda G., Maraschi L., \& Celotti A., 2009b, MNRAS, 402, 497
\bibitem[\protect\citeauthoryear{Ghisellini \& Tavecchio}{2009c}]{b35} Ghisellini G., \& Tavecchio F., 2009c, MNRAS, 397, 985
\bibitem[\protect\citeauthoryear{Ghisellini et al.}{2010}]{b37} Ghisellini G., Tavecchio F., Foschini L., Ghirlanda G., Maraschi L., \& Celotti A., 2010, MNRAS, 402, 497
\bibitem[\protect\citeauthoryear{Ghisellini et al.}{2011}]{b38} Ghisellini G., Tavecchio F., Foschini L., \& Ghirlanda G., 2011, MNRAS, 414, 2674
\bibitem[\protect\citeauthoryear{Ghisellini et al.}{2011}]{b38} Giroletti M., Giovannini G., Taylor G.B., \& Falomo R., 2004a, ApJ, 613, 752
\bibitem[\protect\citeauthoryear{Ghisellini et al.}{2011}]{b38} Giroletti M. et al., 2004b, ApJ, 600, 127
\bibitem[\protect\citeauthoryear{Gu et al.}{2001}]{b39} Gu M., Cao X., \& Jiang D.R., 2001, MNRAS, 327, 1111
\bibitem[\protect\citeauthoryear{Gu et al.}{2009}]{b39} Gu M., Cao X., \& Jiang D.R., 2009, MNRAS, 396, 984
\bibitem[\protect\citeauthoryear{Gu et al.}{2009}]{b39} Heinz S., \& Sunyaev R.A., 2003, MNRAS, 343, 59
\bibitem[\protect\citeauthoryear{Gu et al.}{2009}]{b39} Henri G., \& Sauge L., 2006, ApJ, 640, 185
\bibitem[\protect\citeauthoryear{Gu et al.}{2009}]{b39} Hovatta T., Valtaoja E., Tornikoski M., \& Lahteenmaki A., 2009, A\&A, 494, 527
\bibitem[\protect\citeauthoryear{Jorstad et al.}{2005}]{b39} Jorstad S.G. et al., 2005, ApJ, 130, 1418
\bibitem[\protect\citeauthoryear{Jorstad et al.}{2005}]{b39} Kovalev Y.Y. et al., 2005, AJ, 130, 2473
\bibitem[\protect\citeauthoryear{Leon¨CTavares et al.}{2011a}]{b42} Leon-Tavares J., Valtaoja E., Chavushyan V.H. et al., 2011a, MNRAS, 411, 1127
\bibitem[\protect\citeauthoryear{Leon¨CTavares et al.}{2011b}]{b42} Leon-Tavares J. et al., 2011b, A\&A, 532, 146
\bibitem[\protect\citeauthoryear{Liu et al.}{2006}]{b42} Liu Y., Jiang D.R., \& Gu M.F., 2006, ApJ, 637, 669
\bibitem[\protect\citeauthoryear{Machalski \& Jamrozy}{2006}]{b43} Machalski J., \& Jamrozy M., 2006, A\&A, 454, 95
\bibitem[\protect\citeauthoryear{Mannheim \& Biermann}{1992}]{b44} Mannheim K., \& Biermann P.L., 1992, A\&A, 253, L21
\bibitem[\protect\citeauthoryear{Mannheim}{1993}]{b45} Mannheim K., 1993, A\&A, 269, 67
\bibitem[\protect\citeauthoryear{Maraschi et al.}{1992}]{b46} Maraschi L., Ghisellini G., \& Celotti A., 1992, ApJ, 397, L5
\bibitem[\protect\citeauthoryear{Maraschi et al.}{1992}]{b46} Maraschi L., \& Tavecchio F., 2003, ApJ, 593, 667
\bibitem[\protect\citeauthoryear{Mazin \& Raue}{2007}]{b47} Mazin D., Raue M., 2007, A\&A, 471, 439
\bibitem[\protect\citeauthoryear{Meier et al.}{2001}]{b48} Meier D., Koide S., \& Uchida Y., 2001, Science, 291, 84
\bibitem[\protect\citeauthoryear{Meier et al.}{2001}]{b48} Merloni A. et al., 2003, MNRAS, 345, 1057
\bibitem[\protect\citeauthoryear{Meyer et al.}{2011}]{b48} Meyer E.T., Fossati G., Georganopoulos M., \& Lister M.L., 2011, ApJ, 740, 98
\bibitem[\protect\citeauthoryear{Nemmen et al.}{2001}]{b48} Nemmen R.S., Georganopoulos M., Guiriec S., Meyer E.T., Gehrels N., \& Sambruna R.M., 2012, Science, 338, 1445
\bibitem[\protect\citeauthoryear{Netzer et al.}{1990}]{b48} Netzer H., 1990, Active Galactic Nuclei, 57
\bibitem[\protect\citeauthoryear{Padovani}{1992}]{b49} Padovani P., 1992, A\&A, 256, 399
\bibitem[\protect\citeauthoryear{Padovani}{1992}]{b49} Piner B.G., Pant N., \& Edwards P.G., 2005, ApJ, 622, 168
\bibitem[\protect\citeauthoryear{Padovani}{1992}]{b49} Piner B.G., Pant N., \& Edwards P.G., 2008, ApJ, 678, 64
\bibitem[\protect\citeauthoryear{Padovani}{1992}]{b49} Piner B.G., Pant N., \& Edwards P.G., 2010, ApJ, 723, 1150
\bibitem[\protect\citeauthoryear{Padovani}{1992}]{b49} Piner B.G., \& Edwards P.G., 2013, EPJ Web of Conferences, 6104021P
\bibitem[\protect\citeauthoryear{Pushkarev}{2009}]{b49} Pushkarev A.B., Kovalev Y.Y., Lister M.L., \& Savolainen T., 2009, A\&A, 507, L33
\bibitem[\protect\citeauthoryear{Rawlings \& Saunders}{1991}]{b50} Rawlings S., \& Saunders R., 1991, Nature, 349, 138
\bibitem[\protect\citeauthoryear{Sambruna et al.}{2006}]{b51} Sambruna R.M., Gliozzi M., Tavecchio F. et al., 2006, ApJ, 652, 146
\bibitem[\protect\citeauthoryear{Sbarrato et al.}{2012}]{b52} Sbarrato T., Ghisellini G., Maraschi L., \& Colpi M., 2012, MNRAS, 421, 1764
\bibitem[\protect\citeauthoryear{Scarpa \& Falomo}{1997}]{b53} Scarpa R., \& Falomo R., 1997, A\&A, 325, 109
\bibitem[\protect\citeauthoryear{Scarpa \& Falomo}{1997}]{b53} Senturk G.D., Errando M., Bottcher M., \& Mukherjee R., 2013, ApJ, 764, 119
\bibitem[\protect\citeauthoryear{Serjeant et al.}{1998}]{b54} Serjeant S., Rawlings S., \& Maddox S.J., 1998, MNRAS, 294, 494
\bibitem[\protect\citeauthoryear{Serjeant et al.}{1998}]{b54} Shaw M.S., Romani R.W., Cotter G. et al., 2012, ApJ, 748, 49
\bibitem[\protect\citeauthoryear{Serjeant et al.}{1998}]{b54} Shen Y., Richards G.T., Strauss M.A. et al., 2011, ApJS, 194, 45
\bibitem[\protect\citeauthoryear{Sikora et al.}{1994}]{b56} Sikora M., Begelman M.C., \& Rees M.J., 1994, ApJ, 421, 153
\bibitem[\protect\citeauthoryear{Stanev \& Franceschini}{1999}]{b56} Stanev T., \& Franceschini A., 1998, ApJ, 494, L159
\bibitem[\protect\citeauthoryear{Stecker et al.}{1992}]{b56} Stecker F.W., de Jager O.C., \& Salamon M.H., 1992, ApJ, 390, L49
\bibitem[\protect\citeauthoryear{Tavecchio et al.}{2000}]{b57} Tavecchio F., Maraschi L., Sambruna R.M. \& Urry C.M., 2000, ApJ, 544, L23
\bibitem[\protect\citeauthoryear{Tavecchio et al.}{2004}]{b58} Tavecchio F., Maraschi L., Sambruna R.M., Urry C.M., Cheung C.C., Gambill J.K., \& Scarpa R., 2004, ApJ, 614, 64
\bibitem[\protect\citeauthoryear{Tavecchio et al.}{2007}]{b59} Tavecchio F., Maraschi L., Ghisellini G., Kataoka J., Foschini L., Sambruna R.M. \& Tagliaferri G., 2007, ApJ, 665, 980
\bibitem[\protect\citeauthoryear{Tavecchio et al.}{2011}]{b57} Tavecchio F., Ghisellini G., Ghirlanda G., Foschini L., \& Maraschi L., 2010, MNRAS, 401, 1570
\bibitem[\protect\citeauthoryear{Tavecchio et al.}{2011}]{b57} Tavecchio F., Ghisellini G., Bonnoli G., \& Foschini L., 2011, MNRAS, 414, 3566
\bibitem[\protect\citeauthoryear{Urry \& Padovani}{1995}]{b60} Urry C.M., \& Padovani P., 1995, PASP, 107, 803
\bibitem[\protect\citeauthoryear{Urry \& Padovani}{1995}]{b60} Vestergaard M., \& Peterson B.M., 2006, ApJ, 641, 689
\bibitem[\protect\citeauthoryear{Urry \& Padovani}{1995}]{b60} Vestergaard M., \& Osmer P.S., 2009, ApJ, 699, 800
\bibitem[\protect\citeauthoryear{Volonteri}{1995}]{b60} Volonteri M., Sikora M., \& Lasota J.P., 2007, ApJ, 667, 704
\bibitem[\protect\citeauthoryear{Wagner}{1995}]{b60} Wagner R., 2007, in ASP Conf. Ser., Extragalactic Jets: Theory and Observation from Radio to Gamma Ray, ed. T. A. Rector \& D. S. De Young, (San Francisco: ASP), in press (arXiv:0706.4439)
\bibitem[\protect\citeauthoryear{Wang et al.}{2004}]{b62} Wang J.M., Luo B., \& Ho L.C., 2004, ApJ, 615, L9
\bibitem[\protect\citeauthoryear{Woo \& Urry}{2002}]{b63} Woo J.H., \& Urry C.M., 2002, ApJ, 579, 530
\bibitem[\protect\citeauthoryear{Xie et al.}{1991}]{b64} Xie G.Z. et al., 1991, A\&A, 249, 65
\bibitem[\protect\citeauthoryear{Xie et al.}{1997}]{b65} Xie G.Z., Zhang Y.H., \& Fan J.H., 1997, ApJ, 477, 114
\bibitem[\protect\citeauthoryear{Xie et al.}{2004}]{b66} Xie G.Z., Zhou S.B., \& Liang E.W., 2004, AJ, 127, 53
\bibitem[\protect\citeauthoryear{Xie et al.}{2006}]{b67} Xie G.Z. et al., 2006, AJ, 131, 1210
\bibitem[\protect\citeauthoryear{Xie et al.}{2007}]{b68} Xie G.Z., Dai H., \& Zhou S.B., 2007, AJ, 134, 1464
\bibitem[\protect\citeauthoryear{Zhang et al.}{2012}]{b69} Zhang J., Liang E.W., Zhang S.N., \& Bai J.M., 2012, ApJ, 752, 157
\bibitem[\protect\citeauthoryear{Zhang et al.}{1997}]{b70} Zhang L., \& Cheng K.S., 1997, ApJ, 488, 94
\bibitem[\protect\citeauthoryear{Zhou \& Cao}{2009}]{b71} Zhou M., \& Cao X., 2009, RAA, 9, 293
\end{thebibliography}
\end{document}